\documentclass[lettersize,journal]{IEEEtran}

\usepackage[english]{babel}
\usepackage{glossaries}
\usepackage{subfig}
\usepackage{braket}
\usepackage{nicefrac}
\usepackage{multirow}
\usepackage{siunitx}
\usepackage{booktabs}
\usepackage{hyperref}
\usepackage{layouts}
\usepackage{arydshln}
\usepackage{balance}
\usepackage{orcidlink}
\usepackage{soul}
\usepackage{amsmath, amssymb}
\usetikzlibrary{decorations.pathmorphing, shapes.geometric}

\newacronym{gnn}{GNN}{Graph Neural Network}
\newacronym{ttl}{TTL}{Time-to-Live}

\newcommand{\TM}[1]{} 
\newcommand{\AL}[1]{} 

\usepackage{todonotes}

\begin{document}
\bstctlcite{IEEEexample:BSTcontrol}

\title{RELiQ: Scalable Entanglement Routing via Reinforcement Learning in Quantum Networks}

\author{
\IEEEauthorblockN{Tobias Meuser\IEEEauthorrefmark{1}\orcidlink{0000-0002-2008-5932}, Jannis Weil\IEEEauthorrefmark{1}\IEEEauthorrefmark{2}\orcidlink{0000-0001-5439-9131}, Aninda Lahiri\IEEEauthorrefmark{3}\orcidlink{0000-0003-4833-8046}, Marius Paraschiv\IEEEauthorrefmark{3}\orcidlink{0000-0002-4761-9842}}\\
\IEEEauthorblockA{
\IEEEauthorrefmark{1}Communication Networks Lab, Technical University of Darmstadt, tobias.meuser@kom.tu-darmstadt.de\\
\IEEEauthorrefmark{2}Institute of Communication Technology, Leibniz University Hannover, jannis.weil@ikt.uni-hannover.de\\
\IEEEauthorrefmark{3}Quantum Information Group, IMDEA Networks, \{aninda.lahiri, marius.paraschiv\}@imdea.org
}
\vspace{-0.45cm}
}

\maketitle

\begin{abstract}
Quantum networks are becoming increasingly important because of advancements in quantum computing and quantum sensing, such as recent developments in distributed quantum computing and federated quantum machine learning. Routing entanglement in quantum networks poses several fundamental as well as technical challenges, including the high dynamicity of quantum network links and the probabilistic nature of quantum operations.
Consequently, designing hand-crafted heuristics is difficult and often leads to suboptimal performance, especially if global network topology information is unavailable.

In this paper, we propose RELiQ, a reinforcement learning-based approach to entanglement routing that only relies on local information and iterative message exchange.
Utilizing a graph neural network, RELiQ learns graph representations and avoids overfitting to specific network topologies -- a prevalent issue for learning-based approaches.
Our approach, trained on random graphs, consistently outperforms existing local information heuristics and learning-based approaches when applied to random and real-world topologies. When compared to global information heuristics, our method achieves similar or superior performance because of its rapid response to topology changes.

\end{abstract}


\section{Introduction} 
\label{introduction}

\glsresetall

Quantum computers promise to revolutionize a wide array of fields by solving critical problems in areas ranging from security and encryption to physical system modeling and quantum chemistry. Recent advances have transitioned quantum computing from theoretical constructs to practical implementations \cite{preskill, app1, app2}, with several platforms now offering quantum hardware via a compute-as-a-service model. Despite still being in an early stage of development, many applications such as Blind Quantum Computation \cite{bqc1, bqc2}, Federated Quantum Machine Learning \cite{fqml1, fqml2, fqml3}, or Distributed Quantum Computing \cite{dqc1, dqc2} have emerged, which require quantum computers and quantum sensors to exchange states via quantum teleportation. This process necessitates prior entanglement sharing between nodes, underscoring the need for robust quantum networks.

Quantum networks can be viewed as an additional layer over a classical fiber-based network, in which \emph{elementary links}, represented by entangled pairs, are established between neighboring nodes. The entanglement is then transmitted across the network using quantum repeaters \cite{qnet_review}. Research in quantum networks, designed to facilitate the transfer of entangled quantum states over extensive distances and among multiple nodes, has gained significant momentum. These works address complex issues such as entanglement routing \cite{mprouting1, mprouting2, mprouting3, subgr_comp2}, repeater design \cite{repeater}, and quantum memories \cite{memory1, memory2}. The primary goal of a quantum network is to ensure the efficient distribution of entanglements across communication nodes. This process is constrained by the fact that quantum states cannot be cloned, rendering traditional signal amplification methods inapplicable. Instead, quantum networks employ \emph{entanglement swapping} \cite{ent_swap} to extend the reach of the entanglement distribution beyond the direct transfer limit~\cite{distrib_length}.

Entanglement distribution is further complicated by the necessity to store entangled qubits in quantum memories, where \textit{state fidelities} deteriorate over time as the qubits decay \cite{memory1, memory2}.
This degradation is exacerbated by entanglement swapping, which also diminishes the fidelity of the created end-to-end entanglement. The myriad technical imperfections in network implementation make designing hand-crafted heuristics under these given aconstraints extremely challenging. In addition to information about the underlying network topology, knowledge of the highly dynamic quantum network topology (based on the available elementary links) is also required for optimal routing. Existing heuristics either rely on global information about the quantum network or achieve suboptimal results. Global-information heuristics often ignore the latency induced by monitoring the quantum topology, leading them to use nonexistent elementary links or ignore newly created ones.

To address these issues, we leverage multi-agent reinforcement learning in conjunction with a recurrent message-passing framework, which only utilizes local node-level information to perform entanglement routing. Our RELiQ framework facilitates the creation of global graph representations utilizing a \gls*{gnn} by allowing quantum repeaters to iteratively exchange messages with their direct neighbors. This enhances decision-making capabilities by combining local observations into comprehensive global information, with high resolution for local information and low resolution for distant information. RELiQ relies only on local information available at each repeater and does not require information on the underlying physical network topology or quantum network connectivity. In contrast to other learning-based approaches, our approach significantly improves the generalization and adaptability of routing strategies across diverse physical topologies and dynamic quantum network topologies without necessitating retraining. 

\begin{figure}[t]
    \centering
    \subfloat[Random graph]{\includegraphics[height=3.5cm]{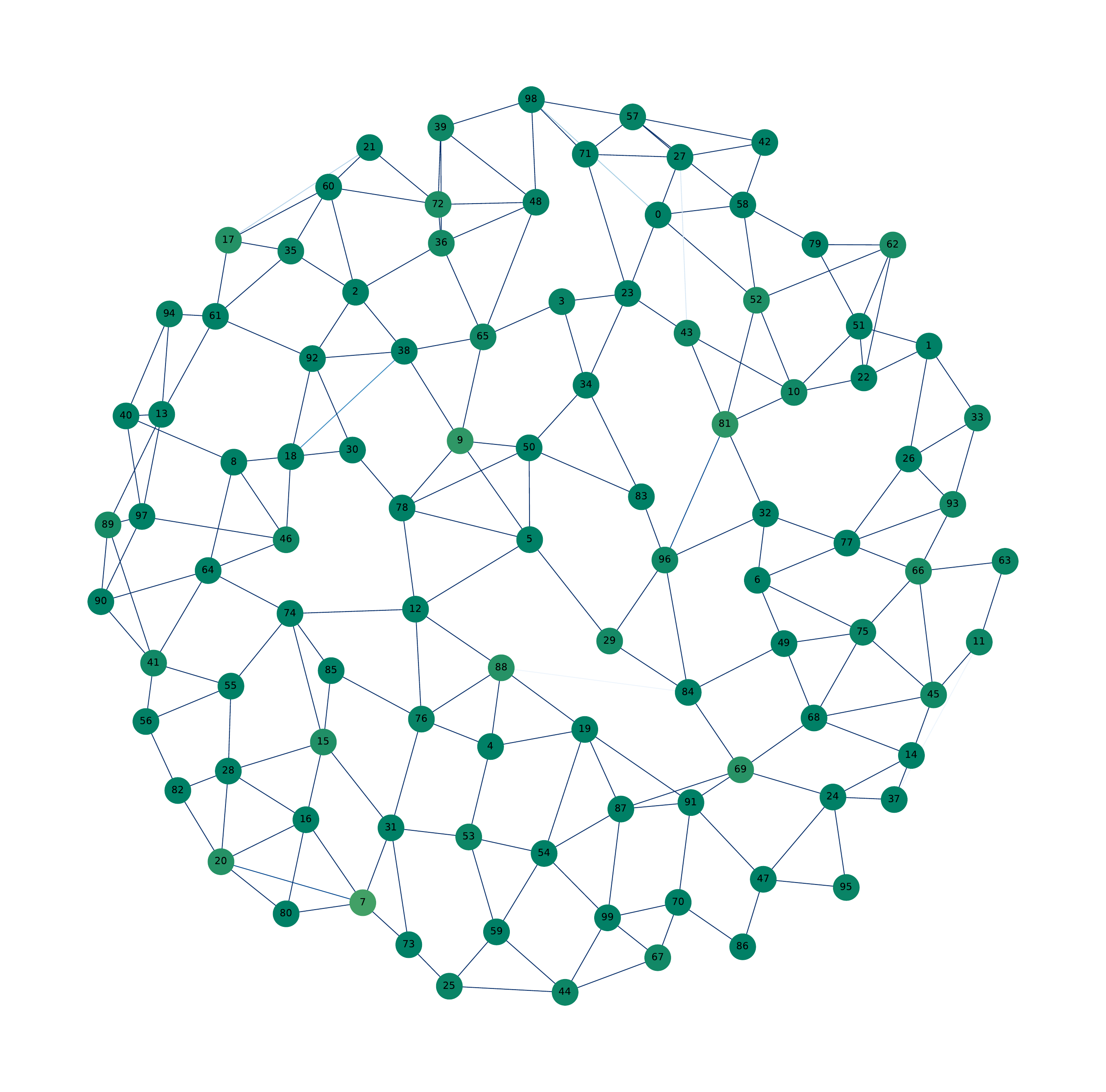}}
    \subfloat[Germany~\cite{sdnlib}]{\includegraphics[height=3.5cm]{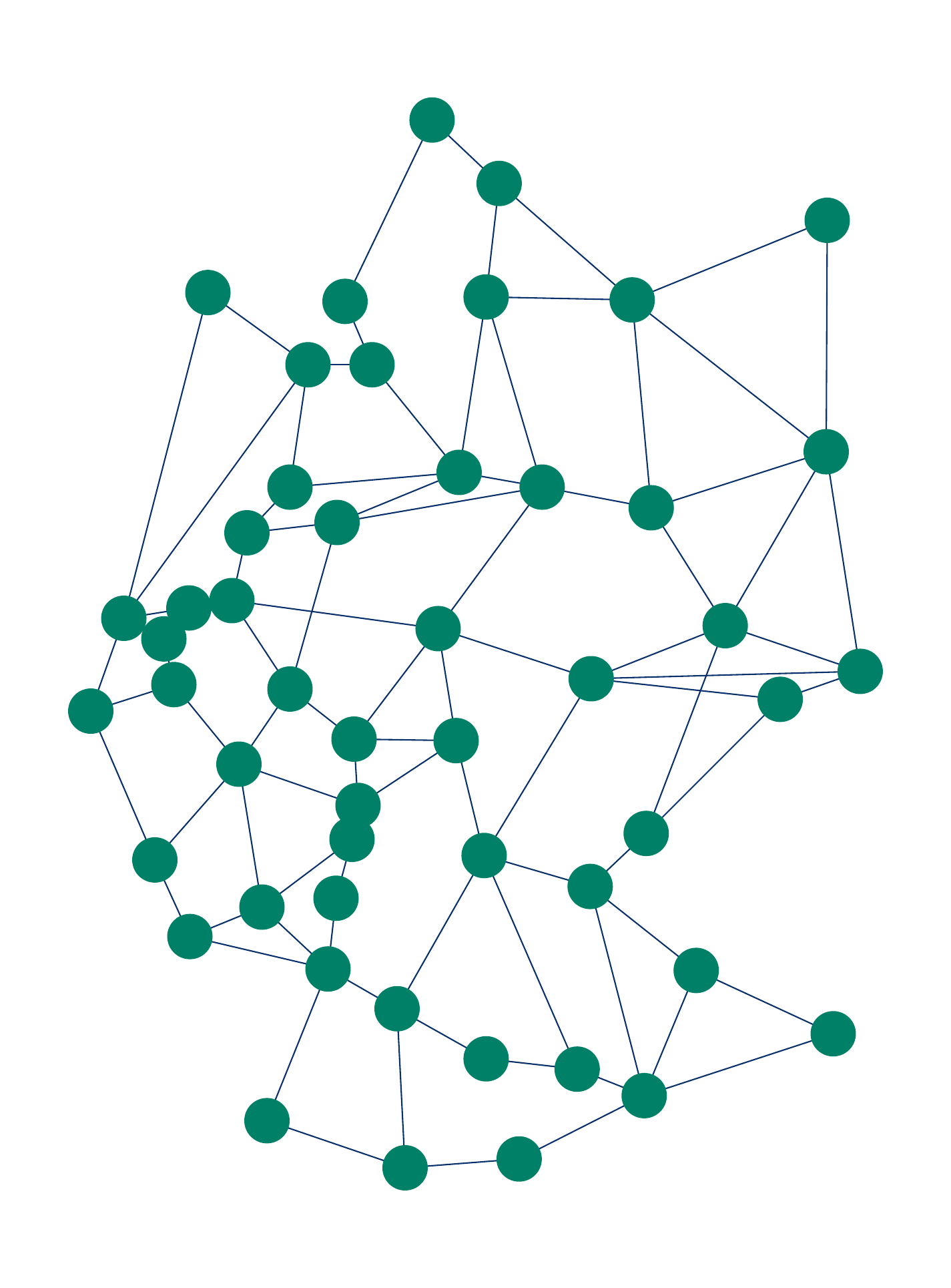}}
    \subfloat[York~\cite{topozoo}]{\includegraphics[height=3.5cm]{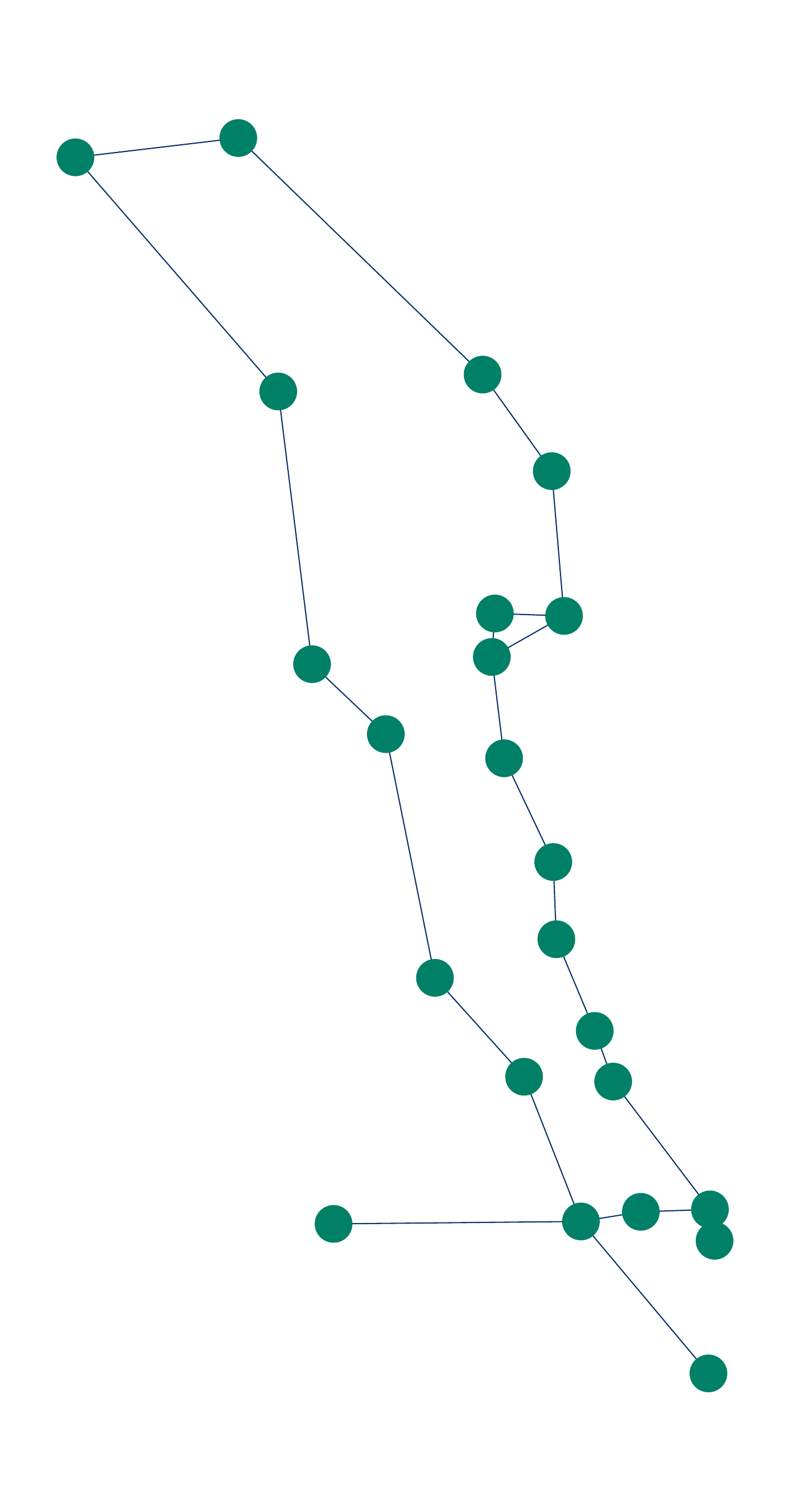}}
    \caption{Sample topologies used for training and evaluation.}
    \label{fig:training_graphs}
\end{figure}

Our work brings three important contributions:
\begin{itemize}
\item We present a framework for local information-based entanglement routing using MARL that surpasses existing local information heuristics and compares favorably with global information heuristics in both random and real-world network topologies, as shown in Fig. \ref{fig:training_graphs}.
\item We extend an existing framework for MARL on graphs to generalize to topologies with varying node count and node degrees, allowing for its application to real-world communication networks.
\item We provide a comparison with three machine learning-based approaches and six heuristics on random graphs and real-world networks, demonstrating the superior performance of our approach across varying topologies, quantum repeater properties, and entanglement qualities.
\end{itemize}

The paper is structured as follows:
In Sec.~\ref{relatedWorks}, we give an overview of related work on entanglement distribution in quantum networks.
Sec.~\ref{methodology} contains a detailed discussion of the methods and techniques employed, with a brief introduction to the concepts of quantum information theory and a formal definition of the quantum network model.
This model is used in Sec.~\ref{rlmodel} to describe the reinforcement learning algorithm and the modeling of the environment.
This is followed by Sec.~\ref{results}, which presents the results of the paper, which are discussed in more detail in Sec.~\ref{discussion}.
The paper concludes with Sec.~\ref{conclusion}, where we summarize our findings and discuss possible future research directions.

\section{Related works}
\label{relatedWorks}

\glsresetall

The field of quantum networks encompasses a broad research area, both theoretical and experimental, aimed at enhancing algorithms and protocols as well as technological infrastructure and capabilities. For a comprehensive overview of recent developments and progress in quantum networking, interested readers are directed to Refs.~\cite{review1, review2, review3, review4, review5, review6}, which provide detailed surveys on the topic.

From a theoretical perspective, significant progress has been made in the area of entanglement routing protocols. Research on bipartite entanglement routing is discussed in Refs.~\cite{routing1, routing2, mpath3, routing4, routing5}. Furthermore, the exploration of multipartite entanglement routing in various network topologies has been addressed in Refs. \cite{mprouting1, mprouting2, mprouting3, mprouting4}.

Certain techniques have been proposed to alleviate the demands on quantum memory by creating virtual graphs based on entangled links, which can be generated as required \cite{vgraph1, vgraph2}. Another approach is multi-path routing \cite{mpath1, mpath2, mpath3, mpath4}, in which multiple end-to-end links between the nodes of interest are created and entanglement distillation is performed to further increase the fidelity of the final state \cite{purification}. A combination of multi-path routing and time-multiplexed quantum repeaters is presented in Ref.~\cite{mpath_multiplex}, while resource under-utilization that leads to the loss of entangled pairs is analyzed in Ref.~\cite{storagenodes}.

Other approaches to entanglement distribution rely on graph-based methods, such as subgraph complementation operations \cite{mprouting4}, which are particularly useful for the study of graph states \cite{subgr_comp2}, where local complementations allow one to establish classes of locally equivalent graphs. In simple topologies, such as linear repeater chains, researchers can determine the optimal entanglement routing solution exactly \cite{repeater_chain1, repeater_chain2}. 

Simultaneously maximizing state fidelity and entanglement distribution rate is fundamental in obtaining useful end-to-end states. This problem has been addressed in \cite{fidelity1, fidelity2} and also using a proposed asynchronous model in \cite{fidelity3}. Finally, the effect of the swapping order on the fidelity of the final state is discussed in \cite{swaporder}, which is an important aspect for creating high-fidelity entanglements.

In this work, we leverage reinforcement learning for quantum entanglement routing.
The idea of using reinforcement learning for path-finding in classical networks has been investigated for a long time~\cite{NIPS1993_4ea06fbc}.
However, as many of these approaches assume centralized control~\cite{ALMASAN2022184}, their scalability is very limited.
In contrast, decentralized approaches tend to offer better scaling; however, the limited observability of the network provides an obstacle to their performance~\cite{9546526}.
This is especially true in dynamic networks like quantum networks, where frequent changes in the quantum topology require an up-to-date view for optimal decisions.
To enable generalizability in path-finding scenarios, \glspl{gnn} have been considered in several works~\cite{JIANG202240,9109574}.
More recently, techniques have been developed that enable generalizability for path-finding utilizing \glspl{gnn}~\cite{10.1145/3229607.3229610,weil2024generalizability}.
One approach focuses on multi-agent reinforcement learning with stateful \glspl{gnn}~\cite{weil2024generalizability}, which allows for a distributed execution and is especially suited for path-finding in quantum networks due to the complexity of exact solutions.
However, the generalizability of this approach is limited to networks with fixed node degree and thus has limited adaptability to changes in the graph, which is an essential aspect for efficient entanglement routing.

In this work, we address the gap in generalizing over graphs with varying node degree and dynamic links using multi-agent reinforcement learning with decentralized execution to enable path-finding in quantum networks.


\section{Methodology}
\label{methodology}

\glsresetall


In this section, we provide a detailed overview of our methodology, starting with an introduction to the basic elements of quantum theory and a description of the quantum network model used in this work.

\begin{figure*}
    \centering
    \subfloat[The upper part of the image depicts the situation before a swap, while the lower part shows the result of a successful swap. The three quantum network nodes are represented by the large gray circles, containing either one or two qubits each, represented by red spheres. Entangled qubits are depicted as blue dashed arches.\label{swapfig}]{\includegraphics[height=5cm, width=6cm]{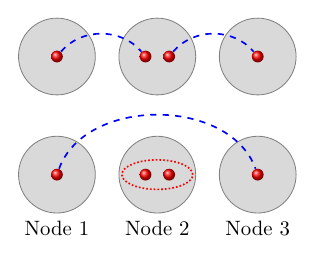}}\quad
    \subfloat[Noisy swap circuit used to simulate entanglement swapping at network nodes. Entangled pairs are established between nodes 1 -- 2 and 2 -- 3, depicted here for representation purposes through two entangling operations. In a real scenario, nodes 1 and 3 create the Bell pairs locally and each send one entangled photon to node 2. Thus, at no point are quantum gates applied across physically distant nodes. The photon transfer operation is noisy and simulated by the addition of two depolarization channels as a next step. The repeater node 2 then performs the Bell State Measurement (BSM) on its two qubits and communicates the classical outcomes ($c_0$ and $c_1$). Node 3 applies conditional Pauli corrections: a Z gate is applied if and only if the measurement outcome $c_0$ is 1, and an X gate is applied if and only if $c_1$ is 1. This protocol establishes a final entangled state between the remote qubits of node 1 and node 3. \label{fig:Qiskit_Circuit}]{\includegraphics[height=5cm, width=11cm]{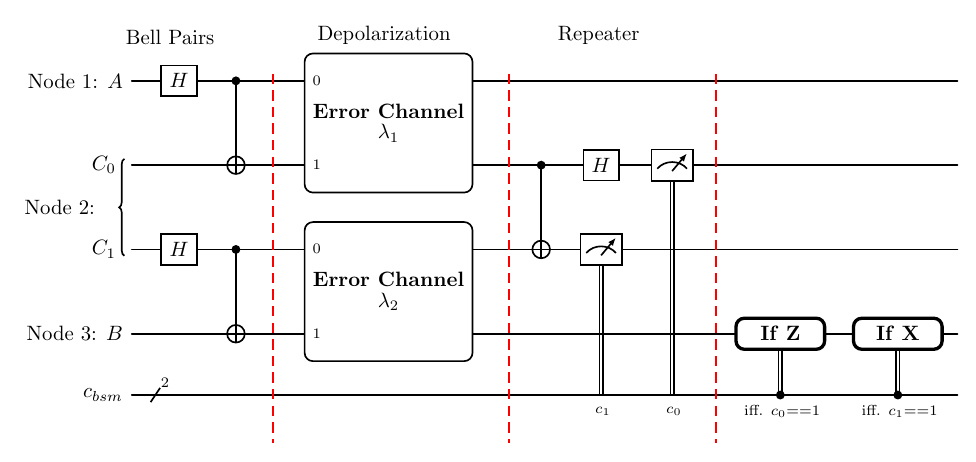}}
    \caption{Illustration of entanglement swapping. Bell pairs are established between neighboring nodes (nodes 1 and 2, nodes 2 and 3). The intermediary node 2, containing two qubits, will play the role of a \textit{quantum repeater} and perform the Bell State Measurement (BSM) on its qubits. The result is an entangled state between the remote qubits of node 1 and node 3.}
\end{figure*}

\subsection{Elements of quantum theory}
\label{elementsofqt}



\glsresetall

This section lays out the fundamental quantum mechanical principles necessary for the subsequent analysis of quantum networks. Core concepts include the definition of a qubit, the nature of superposition and entanglement in multi-qubit systems, with a specific focus on Bell states which serve as the elementary links within these networks. We also introduce a formalism for describing noise in quantum systems using density operators, quantifying state degradation through fidelity, and modeling the decay of Bell pairs into Werner states.
Due to the physical limitations of direct qubit transmission, the creation of long-range entanglements in quantum networks relies on the entanglement swapping protocol. In this work, we simulate this protocol using a noisy circuit model.
These core concepts will then be used throughout the rest of the paper.

A quantum bit, commonly known as a qubit, serves as the fundamental unit in quantum computing, analogous to the classical bit in conventional computing. A qubit possesses the ability to exist not only in discrete states, typically represented as $\ket{0}$ and $\ket{1}$, but also in any superposition of these states. This superposition is expressed as $\ket{\psi} = \alpha_1 \ket{0} + \alpha_2 \ket{1}$, where $\alpha_1$ and $\alpha_2$ are complex coefficients that satisfy the normalization condition $|\alpha_1|^2 + |\alpha_2|^2 = 1$.

When extended to systems composed of multiple qubits, the overall state of the system is represented within the tensor product space of the individual qubit Hilbert spaces. For a two-qubit system, the state may be characterized as either separable or entangled. A separable state implies that the state of the composite system can be described as the tensor product of its constituents, $\ket{\psi_{AB}} = \ket{\psi_A} \otimes \ket{\psi_B}$. Conversely, an entangled state cannot be expressed as such a product, indicating non-classical correlations between the qubits.

In addition to pure states, which are representable as vectors in Hilbert space, quantum systems may also exist in mixed states. Mixed states, which represent statistical mixtures of pure states, are described by a density operator. A pure state $\ket{\psi}$ can be expressed as a density operator by the outer product $\rho = \ket{\psi}\bra{\psi}$.

Important examples of entangled states in two-qubit systems are Bell states, which form the basis for many protocols in quantum information theory. The Bell states, or Bell pairs, are defined as $\ket{\Phi^\pm} = \frac{1}{\sqrt{2}}(\ket{00} \pm \ket{11})$ and $\ket{\Psi^\pm} = \frac{1}{\sqrt{2}}(\ket{01} \pm \ket{10})$. Throughout this paper, we shall refer to systems in a Bell state as forming a \textit{Bell pair} and, in the context of quantum networks, as an \textit{elementary link}.

In the context of quantum communication, the transmission of qubits through a non-ideal channel introduces noise, transitioning a pure state into a mixed state. To quantify the extent to which the transmitted state deviates from the original, the \textit{fidelity} metric is employed. Fidelity measures the amount of state overlap and is defined for a pure state $\ket{\Psi^{+}}$ and a resulting mixed state $\rho$ as $F = \bra{\Psi^{+}}\rho\ket{\Psi^{+}}$~\cite{Nielsen_Chuang_2010}.

The noise effects of the communication channel on a pure state $\ket{\Psi^{+}}$ can be modeled as a transformation into a mixed Werner state $W_F$, as described by \cite{werner_state}:

\begin{equation}
\begin{split}
   W_F = F \ket{\Phi^{+}} \bra{\Phi^{+}} + \\
   & \hspace{-55pt}\frac{1 - F}{3} \left( \ket{\Phi^{-}} \bra{\Phi^{-}} + \ket{\Psi^{+}} \bra{\Psi^{+}} + \ket{\Psi^{-}} \bra{\Psi^{-}} \right)
\end{split}
\label{wernerstate}
\end{equation}

In this representation, the first addend of Eq.~\ref{wernerstate} refers to the component of the initial pure state, while the second addend embodies the isotropic noise introduced by the channel. 

In the context of quantum networks, the establishment of elementary links starts with one node creating an entangled Bell pair locally, which is then transmitted through an optical fiber to its neighbor. The transmission process as well as storage in memory cause the entangled state to decay. Thus, when the elementary link is formed, it is described by the Werner state from Eq.~\ref{wernerstate}, whose fidelity decreases with time.

The direct transmission of entangled photons over optical fibers is severely limited by the photon absorption of the physical channel. Coupled with the impossibility to clone a quantum state and thus to amplify a quantum signal, a novel protocol is required to establish long-distance entangled links. Most entanglement-assisted quantum networks, referred to in the rest of this paper simply as quantum networks, rely on a protocol named \textit{entanglement swapping} \cite{ent_swap, swaporder} to surpass the distance limitations of direct transmission. 

A high-level overview of the swap operation is depicted in Fig.~\mbox{\ref{swapfig}}, a swap at node $2$ creates an end-to-end entanglement between nodes $1$ and $3$. We have implemented the swap operation as a 4-qubit entanglement scheme using Qiskit \mbox{\cite{qiskit2024}}, as shown in Fig.~\mbox{\ref{fig:Qiskit_Circuit}}. The two Bell pairs created at nodes $1$ and $3$ act as input states for the swap circuit. In a real-world implementation, the generation of these Bell pairs can be realized by Spontaneous Parametric Down-Conversion (SPDC), where a high-energy photon is split into a pair of lower-energy, entangled photons. From these photon pairs, one of the photons is captured at the intermediary node 2 and the entanglement is transferred to the $C_0$ and $C_1$ qubits. This transmission introduces noise, which is modeled in the circuit by the depolarization error channels $\lambda_1$ and $\lambda_2$. These channels represent the cumulative effect of real-world imperfections, such as fidelity loss due to transmission through optical fibers and quantum memory storage. The core protocol is executed on the repeater. To perform the swap, the repeater applies a joint local operation on the two qubits it holds ($C_1$ and $C_0$). The repeater then measures both of its qubits in the Bell basis (represented by the two measurement gates). This measurement projects the two qubits onto one of the four Bell states. The classical outcomes of the Bell State Measurement (BSM) are transmitted to the end node $3$. Node $3$ applies classical conditional corrections (Pauli Z/Pauli X gates) to its local qubit $B$. Thus, regardless of the BSM outcome, the qubits $A$ and qubit $B$ are projected into a known, shared Bell state. The successful completion of the swapping protocol creates an entangled state between the remote qubits $A$ and $B$. The fidelity of this final entangled state is limited by the noise introduced during transmission through the depolarization channels and the fidelity of the repeater gate ($f_\text{gate}$).

The discussion of the noise model and the choice of input fidelities is detailed in Appendix \ref{appendix_error}.

\subsection{Quantum network model}
\label{qnetmodel}

\glsresetall

We define the quantum network as a graph $G_Q = (V, E_Q)$, where $V$ represents the set of nodes and $E_Q$ the set of elementary links.
Each node $v \in V$ is equipped with $C_v$ qubits, where $C_v$ can vary depending on the node's intended role within the network. The qubits are stored in quantum memories at each node, with coherence times that limit the operational window for network tasks.
The objective is to establish end-to-end entanglements between the source-destination pairs $\{(v^1_s, v^1_d), \dots, (v^N_s, v^N_d)\}$ with $N$ being the number of source-destination pairs.

Each edge $e \in E_P$ of the physical network $G_P = (V, E_P)$ represents an optical fiber connection between neighboring nodes.
One physical connection can correspond to multiple communication channels between nodes, achieved, for example, by multiplexing.
The overall capacity of the edge $C_e$ corresponds to the number of parallel elementary links (Bell pairs) that can be established along the edge $e = \{u, v\}$, which is limited by the number of qubits stored by the endpoint nodes, $C_e \leq \min{(C_{u}, C_{v})}$, where $u$ and $v$ represent the end nodes of the edge $e$.

The fidelity of elementary links decays over time. This decay is small at the beginning, but then accelerates rapidly until the baseline fidelity $F_B=0.5$ is reached~\mbox{\cite{doi:10.1126/science.add9771}}. The speed of this decay can be influenced by decoupling pulses.
By increasing the number of decoupling pulses, the stability of the qubits improves, increasing their lifetime in quantum memory.
The fidelity of these qubits is modeled using a stretched exponential function as shown in Eq.~\mbox{\ref{eq:fidelity_decay}} (derived from \mbox{\cite{doi:10.1126/science.add9771}}).

\begin{equation}\label{eq:fidelity_decay}
    F(t) = \left(F_0 - F_{B}\right) \cdot \exp\left[-\left(\frac{t}{T_{2,n_{\text{dec}}}}\right)^k\right] + F_B
\end{equation}

\begin{figure}[tb]
    \centering
    \input{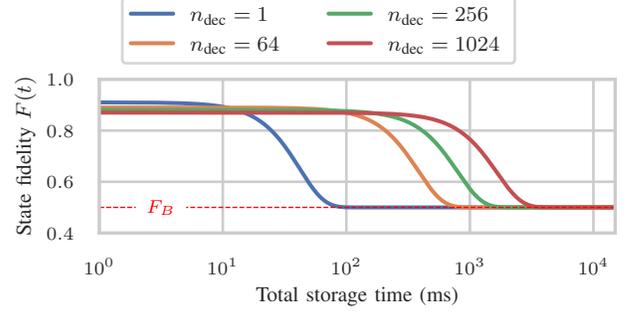}
    \caption{Decay of qubits in quantum memory depending on the number of decoupling pulses $n_{\text{dec}}$, reproduced from~\cite{doi:10.1126/science.add9771}.}
    \label{fig:decay}
\end{figure}

Note that this fidelity describes the noise effects on stored qubits.
$F_0 \leq 1$ is the initial fidelity, and $F_{B} = 0.5$ the baseline fidelity below which the entangled states cannot be used.
$k = 2.2$ is the exponent of the stretch and is chosen to match the decay curve~\mbox{\cite{doi:10.1126/science.add9771}}.
The decoherence time $T_{2,n_{\text{dec}}}$ is determined by the number of decoupling pulses and influences the decoherence speed of the qubits.
With $T_{2,n_{\text{dec}}} \propto n_{\text{dec}}^\beta$~\mbox{\cite{doi:10.1126/science.add9771}}, we can derive $T_{2,n_{\text{dec}}} \approx \mbox{\SI{42}{\milli\second}} \cdot n_{\text{dec}}^\beta$.
The resulting decay function is shown in Fig.~\mbox{\ref{fig:decay}}.
It can be seen that the fidelity remains relatively stable at first, but then decreases rapidly until the baseline fidelity $F_{B}=0.5$ is reached.

\begingroup
\definecolor{tabblue}{RGB}{31, 119, 180}
\definecolor{tabred}{RGB}{214, 39, 40}
\definecolor{tabgreen}{RGB}{44, 160, 44}
\definecolor{taborange}{RGB}{255, 127, 14}

\tikzset{
    node distance=1.5cm,
    node/.style={circle, draw, minimum size=.75cm},
    connection/.style={thick, gray},
    entangled/.style={line width=.5pt, tabblue, decorate, decoration={snake, amplitude=2pt, segment length=7pt}},
    entangled-strong/.style={line width=.5pt, tabblue, decorate, decoration={snake, amplitude=2pt, segment length=7pt}, double, double distance=1pt},
    agent/.style={rectangle, fill=tabred, minimum size=0.3cm},
    path/.style={line width=2.5pt, tabred, opacity=0.4},
    message/.style={->, thick, tabgreen, shorten >=2pt, shorten <=2pt, bend left=15},
    message-label/.style={font=\tiny, tabgreen}
}

\newcommand{\widthfirstphase}{5.65cm}

\begin{figure*}[h]
    \centering
    \begin{tikzpicture}[baseline=(current bounding box.north)]
        \draw[draw=black] (0,0) rectangle ++(\widthfirstphase,-5.3);
        \node at (\widthfirstphase / 2,-0.25) {\textbf{Phase 1}};
        
        \draw[draw=black] (\widthfirstphase,0) rectangle ++(\textwidth - \widthfirstphase,-5.3);
        \node at (\textwidth / 2 + \widthfirstphase / 2,-0.25) {\textbf{Phase 2}};
    \end{tikzpicture}
    
    \vspace{-5cm} 
    
    \begin{minipage}{\textwidth}
        \centering
        \subfloat[Fiber network\label{subfig:fiber}]{
            \begin{tikzpicture}[baseline=0pt]
                \node[node] (A1) at (0,0) {};
                \node[node] (A2) at (0,1.5) {};
                \node[node] (A3) at (0,3) {};
                \node[node] (A4) at (1.2,0.75) {};
                \node[node] (A5) at (1.2,2.25) {};
                
                \draw[connection] (A1) -- (A2);
                \draw[connection] (A2) -- (A3);
                \draw[connection] (A1) -- (A4);
                \draw[connection] (A2) -- (A4);
                \draw[connection] (A2) -- (A5);
                \draw[connection] (A3) -- (A5);
                \draw[connection] (A4) -- (A5);
            \end{tikzpicture}
        }
        \hspace{0.25cm}
        \subfloat[After entanglement generation\label{subfig:entangled}]{
            \begin{tikzpicture}[baseline=0pt]
                \node[node] (A1) at (0,0) {};
                \node[node] (A2) at (0,1.5) {};
                \node[node] (A3) at (0,3) {};
                \node[node] (A4) at (1.2,0.75) {};
                \node[node] (A5) at (1.2,2.25) {};
                
                \draw[entangled-strong] (A1) -- (A2);
                \draw[entangled] (A2) -- (A3);
                \draw[entangled] (A1) -- (A4);
                \draw[connection] (A2) -- (A4);
                \draw[entangled-strong] (A2) -- (A5);
                \draw[entangled] (A3) -- (A5);
                \draw[entangled] (A4) -- (A5);
            \end{tikzpicture}
        }
        \hspace{0.5cm}
        \subfloat[Start of path planning\label{subfig:start}]{
            \begin{tikzpicture}[baseline=0pt]
                \node[node] (A1) at (0,0) {\small $v^1_s$};
                \node[node] (A2) at (0,1.5) {};
                \node[node] (A3) at (0,3) {\small $v^1_d$};
                \node[node] (A4) at (1.2,0.75) {};
                \node[node] (A5) at (1.2,2.25) {};
                
                \draw[entangled-strong] (A1) -- (A2);
                \draw[entangled] (A2) -- (A3);
                \draw[entangled] (A1) -- (A4);
                \draw[connection] (A2) -- (A4);
                \draw[entangled-strong] (A2) -- (A5);
                \draw[entangled] (A3) -- (A5);
                \draw[entangled] (A4) -- (A5);
                
                \node[agent] (PA1) at (-0.5,0) {\small 1};
                
            \end{tikzpicture}
        }
        \hspace{0.25cm}
        \subfloat[After first agent decision\label{subfig:first}]{
            \begin{tikzpicture}[baseline=0pt]
                \node[node] (A1) at (0,0) {\small $v^1_s$};
                \node[node] (A2) at (0,1.5) {};
                \node[node] (A3) at (0,3) {\small $v^1_d$};
                \node[node] (A4) at (1.2,0.75) {};
                \node[node] (A5) at (1.2,2.25) {};
                
                \draw[entangled-strong] (A1) -- (A2);
                \draw[entangled] (A2) -- (A3);
                \draw[entangled] (A1) -- (A4);
                \draw[connection] (A2) -- (A4);
                \draw[entangled-strong] (A2) -- (A5);
                \draw[entangled] (A3) -- (A5);
                \draw[entangled] (A4) -- (A5);
                
                \draw[path] (A1) -- (A2);
                
                \node[agent] (PA2) at (-0.5,1.5) {\small 1};
                
            \end{tikzpicture}
        }
        \hspace{0.25cm}
        \subfloat[Finished path planning\label{subfig:finished}]{
            \begin{tikzpicture}[baseline=0pt]
                \node[node] (A1) at (0,0) {\small $v^1_s$};
                \node[node] (A2) at (0,1.5) {};
                \node[node] (A3) at (0,3) {\small $v^1_d$};
                \node[node] (A4) at (1.2,0.75) {};
                \node[node] (A5) at (1.2,2.25) {};
                
                \draw[entangled-strong] (A1) -- (A2);
                \draw[entangled] (A2) -- (A3);
                \draw[entangled] (A1) -- (A4);
                \draw[connection] (A2) -- (A4);
                \draw[entangled-strong] (A2) -- (A5);
                \draw[entangled] (A3) -- (A5);
                \draw[entangled] (A4) -- (A5);
                
                \draw[path] (A1) -- (A2) -- (A3);
                
                \node[agent] (PA3) at (-0.5,3) {\small 1};
                
            \end{tikzpicture}
        }
        \hspace{0.25cm}
        \subfloat[Agent is reset to $v^1_s$\label{subfig:reset}]{
            \begin{tikzpicture}[baseline=0pt]
                \node[node] (A1) at (0,0) {\small $v^1_s$};
                \node[node] (A2) at (0,1.5) {};
                \node[node] (A3) at (0,3) {\small $v^1_d$};
                \node[node] (A4) at (1.2,0.75) {};
                \node[node] (A5) at (1.2,2.25) {};
                
                \draw[entangled] (A1) -- (A2);
                \draw[connection] (A2) -- (A3);
                \draw[entangled] (A1) -- (A4);
                \draw[connection] (A2) -- (A4);
                \draw[entangled-strong] (A2) -- (A5);
                \draw[entangled] (A3) -- (A5);
                \draw[entangled] (A4) -- (A5);
                
                \node[agent] (PA1) at (-0.5,0) {\small 1};
                
            \end{tikzpicture}
        }
    \end{minipage}
    \caption{Visualization of the phases of our quantum network model. The quantum repeaters are depicted as the nodes of the graph, which are connected via fiber. If elementary links are available between two repeaters, each blue wavy line represents one of these links. The agent is represented as a red rectangle, which moves across the network in Phase 2. The currently planned path is depicted as a red line.}\label{fig:phases}
\end{figure*}
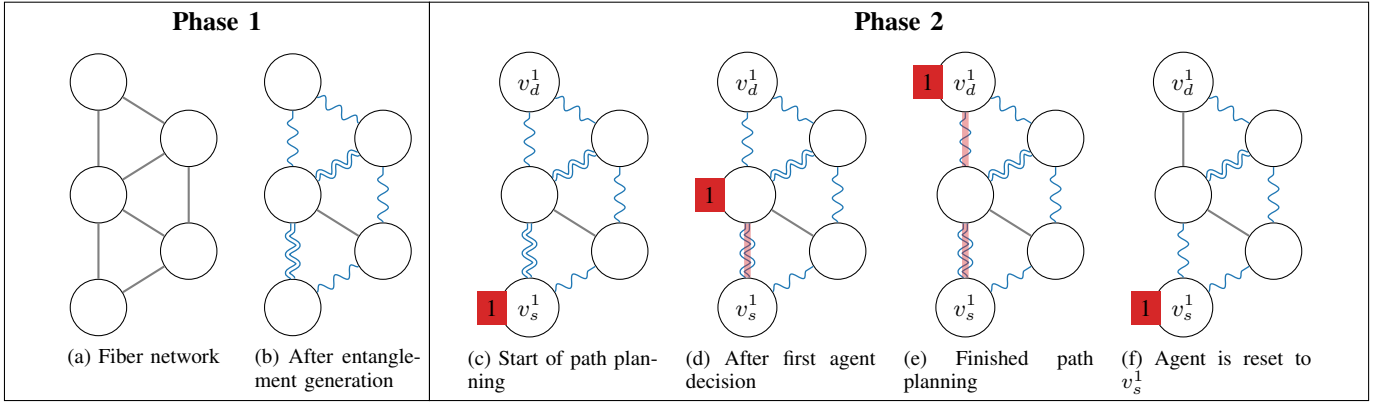
\endgroup

In line with recent literature~\cite{qpath_and_qleap}, we assume two phases for our network model, which are depicted in Fig.~\ref{fig:phases}: 
\begin{itemize}
    \item \textit{Phase 1} - All nodes start without any elementary links as depicted in Fig.~\mbox{\ref{subfig:fiber}} and attempt to establish the maximum number of elementary links with their neighbors. To avoid sparse graphs, we impose the constraint that the elementary links should be distributed as uniformly as possible among neighbors. This is achieved by dividing the available qubits $C_v$ of a node $v$ into $n_v$ buckets, with $n_v$ being the number of neighbors of $v$ in the fiber topology, and reserving the available qubits of each bucket for one neighbor. If the number of qubits is lower than the number of neighbors, the qubits should be distributed such that the overall connectivity of the quantum topology is ensured. This can be achieved by constructing a maximum spanning tree using weights that reflect the available qubits on the repeaters connected through each edge. Any remaining qubits may then be distributed between the remaining neighbors. After this phase, elementary links between most nodes have been established as shown in Fig.~\mbox{\ref{subfig:entangled}}, but long-distance links may lead to neighbors without elementary links. 

    \item \textit{Phase 2} - During the second phase, the creation of elementary links is continued, but end-to-end entanglements between all source-destination pairs $\{(v^1_s, v^1_d), \dots, (v^N_s, v^N_d)\}$ are established through agents that traverse the quantum network. This is essentially a multi-objective optimization problem, as we aim to maximize both end-to-end fidelity of the final links for each source-destination pair, as well as the overall \textit{Entanglement Distribution Rate} (EDR), which represents the total number of final states established per unit time. Each agent iteratively plans the full route first, which is depicted in Fig.~\mbox{\ref{subfig:start}} -- \mbox{\ref{subfig:finished}}. While an agent only plans routes sequentially, the usage of multiple agents for a single source-destination pair is possible to enable multiplexing. Similarly to related work~\mbox{\cite{qpath_and_qleap}}, the agent reserves elementary links along the path during path planning to signal to other agents that these links are no longer available. After a path has been planned, the agent is reset to its initial position as shown in Fig.~\mbox{\ref{subfig:reset}} and the repeaters on the path try to create an end-to-end entanglement. This consumes one elementary link on each fiber along the path and fails if at least one swap operation is unsuccessful or a fiber link without any available elementary links is utilized.
\end{itemize}

An important distinction must be made here between the underlying physical network, formed by optical fiber connections, and the evolving quantum network graph. Although these would coincide in an ideal situation at the end of \textit{Phase~1}, assuming all neighbors successfully created elementary links, the swap operations performed during \textit{Phase~2} will consume elementary links, causing the topologies of the physical network and quantum network to diverge significantly.

For the creation of elementary links, the two connected quantum repeaters exchange entangled photons via a fiber.
The number of elementary links that are created simultaneously depends on the available quantum storage.
Both repeaters create qubits and store one of them, while transmitting the other qubit to the connected repeater.
The success probability of each individual transmitted entangled photon reaching the other node can be estimated using the attenuation constant $\alpha$ of the fiber in $\nicefrac{\text{dB}}{\text{km}}$ and the length of the fiber $L$ in $\text{km}$:

\begin{figure}[b]
   \centering
    \input{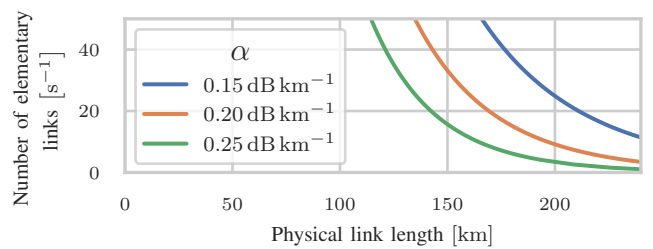}
    \caption{Number of successfully created elementary links per second for different link lengths and attenuation constants $\alpha$. The number of successfully created elementary links decreases drastically with increasing link length.}
    \label{fig:elementary_link_generation}
\end{figure}

\begin{equation}
    p = \exp{\left(-\dfrac{\alpha L}{10}\right)}
\end{equation}

Single-mode fibers typically have an attenuation constant of approx. \mbox{\SI{0.2}{\decibel\per\kilo\meter}}~\mbox{\cite{doi:10.1049/el:19790077}}, while the theoretical minimum is at approx. \mbox{\SI{0.15}{\decibel\per\kilo\meter}}~\mbox{\cite{4542888}}.
Due to the limited available quantum storage, each repeater must wait for feedback from the other repeater to send another entangled photon.
In total, a repeater could transmit and verify up to $n = \lfloor \nicefrac{cT}{2L} \rfloor$ entanglements per time period $T$ and available slot in quantum memory, with $c$ being the speed of light in the fiber.
However, for very short links, this exchange rate might be higher than the generation rate of quantum repeaters, which is often around a few MHz~\cite{RWL+24}, in which case $n$ is reduced to the maximum generation rate available per neighbor.

If we consider the successful creation of two entanglements as independent events, we can use the binomial distribution to calculate the number of entanglements that can be created.
Note that this number decays substantially with the increasing length of the physical link, as shown in Fig.~\ref{fig:elementary_link_generation} for $T=\SI{1}{\second}$ and one qubit available in the storage of the sender.

\section{System design}
\label{rlmodel}

\glsresetall

In this section, the architecture of the reinforcement learning model is presented.
%
Existing works on reinforcement learning for graph-based problems often either train for one specific graph or rely on the structure of the graph, such as link connections and shortest-path information, to guide their decision-making processes.
Thus, they either must be adapted and retrained for each change in the underlying graph, making them unusable for applications in communications and especially quantum networks, or require global information about the graph structure, which is especially hard to obtain for quantum networks.
In contrast, Weil et al.~\cite{weil2024generalizability} created an abstraction layer, a position-dependent graph representation, between the underlying graph and the reinforcement learning agents, ensuring a similar observation for reinforcement learning agents even if the underlying graph has changed.
In their approach, they utilize learned graph representations based on a recurrent \gls*{gnn} to approximate the network topology and use the information obtained in reinforcement learning.
They relied on identifier-based addressing and required a fixed node degree for all nodes in the network, drastically limiting the applicability to real-world networks.
In the following, we extend their architecture to improve its applicability to quantum entanglement routing.

\subsection{System model}
Weil et al.~\cite{weil2024generalizability} differentiate between nodes and agents, with nodes connected through edges to form a graph.
In our model, the nodes $V$ of the graph are the quantum repeaters, which are connected to each other through optical fiber links.
These fiber links represent the edges $E_P$ of the graph.
Each edge $e \in E_P$ has a set of available elementary links $\mathcal{C}_e$, with each link $c \in \mathcal{C}_e$ being characterized by its fidelity $F_{c}$.

As shown in Fig.~\ref{subfig:first} -- \ref{subfig:reset}, the agents are located at the repeater, currently making the decision on the next hop to add to a route and thus symbolizing the current progress of the iterative planning of a route.
Thus, the agent is the decision-making entity for the quantum repeater and is able to access its state information.
Each agent is responsible for creating end-to-end entanglements between a source-destination pair $(v^i_s, v^i_d) \in V \times V$ with $i \in \{1, \ldots, N\}$, where $N$ is the total number of source-destination pairs.

As also visualized in Fig.~\mbox{\ref{subfig:start}}, each agent is placed at $v^i_c = v^i_s$ at the start of the simulation.
They are reset to this node if they either successfully reach the destination $v^i_d$ (Fig.~\mbox{\ref{subfig:finished}}) or are dropped due to exceeding a maximum \gls*{ttl}.
Once the destination has been reached, the repeaters along the path perform the swap operation to create an end-to-end entanglement, starting from the first repeater in the path until the destination has been reached.
Each swap also removes the elementary links from the associated edges.
Even if the destination $v^i_d$ is reached, the routing can be unsuccessful for three reasons: i) there are not enough elementary links available at any link along the route, ii) the swap fails, or iii) the fidelity of the link decreases beyond a chosen threshold $F_{th}$.
In this work, we chose this threshold to be $F_{th}=F_B=0.5$, as below this fidelity, the entanglement can no longer be distilled.
For some applications like quantum key distribution that require higher fidelities to perform properly, either $F_{th}$ should be increased or distillation should be performed.

\begingroup
\definecolor{tabblue}{RGB}{31, 119, 180}
\definecolor{tabred}{RGB}{214, 39, 40}
\definecolor{tabgreen}{RGB}{44, 160, 44}
\definecolor{taborange}{RGB}{255, 127, 14}

\captionsetup[sub]{width=2.2cm}
\tikzset{
    node distance=2.5cm,
    node/.style={circle, draw, minimum size=.75cm},
    connection/.style={thick, gray},
    entangled/.style={line width=.5pt, tabblue, decorate, decoration={snake, amplitude=2pt, segment length=7pt}},
    entangled-strong/.style={line width=.5pt, tabblue, decorate, decoration={snake, amplitude=2pt, segment length=7pt}, double, double distance=1pt},
    agent/.style={rectangle, fill=tabred, minimum size=0.4cm},
    agent-central/.style={regular polygon, regular polygon sides=3, fill=tabred, minimum size=0.5cm, draw=black, line width=1pt},
    path/.style={line width=2.5pt, tabred, opacity=0.4},
    message/.style={->, thick, tabgreen, shorten >=2pt, shorten <=2pt, bend left=15},
    message-label/.style={font=\tiny, tabgreen},
    gnn/.style={regular polygon, regular polygon sides=6, fill=tabgreen, minimum size=0.4cm, inner sep=0pt, rotate=30},
    control/.style={->, line width=2.5pt, taborange, shorten >=2pt, shorten <=2pt}
}

\begin{figure}[t]
    \centering
    
    \begin{minipage}{\linewidth}
        \centering
        
        \subfloat[Relevant related work~\cite{qpath_and_qleap}: All information is collected at a central node (red triangle).\label{subfig:monitoring_centralized}]{
            \begin{tikzpicture}[baseline=0pt]
                \node[node] (A1) at (0,0) {\small $v^1_s$};
                \node[node] (A2) at (0,2) {};
                \node[node] (A3) at (0,4) {\small $v^1_d$};
                \node[node] (A4) at (2,1) {};
                \node[node] (A5) at (2,3) {};
                
                \draw[entangled-strong] (A1) -- (A2);
                \draw[entangled] (A2) -- (A3);
                \draw[entangled] (A1) -- (A4);
                \draw[connection] (A2) -- (A4);
                \draw[entangled-strong] (A2) -- (A5);
                \draw[entangled] (A3) -- (A5);
                \draw[entangled] (A4) -- (A5);
                
                \draw[path] (A1) -- (A2);
                
                \node[agent] (PA2) at (-0.5,2) {\small 1};

                \node[agent-central] (PA4) at (2,1) {};

                \draw[control] (PA4) to[bend left=40] (PA2);

                \draw[message] (A1) to[bend right=20] node[message-label, pos=0.5] {} (A4);
                \draw[message] (A2) to[bend left=15] node[message-label, pos=0.5] {} (A4);
                \draw[message] (A3) to[bend left=15] node[message-label, pos=0.5] {} (A4);
                \draw[message] (A5) to[bend left=15] node[message-label, pos=0.5] {} (A4);
            \end{tikzpicture}
        }
        \hspace{0.25cm}
        \subfloat[Our approach: Information is only shared with direct neighbors and aggregated using a \gls*{gnn} (green polygons).\label{subfig:monitoring_distributed}]{
            \begin{tikzpicture}[baseline=0pt]
                \node[node] (A1) at (0,0) {\small $v^1_s$};
                \node[node] (A2) at (0,2) {};
                \node[node] (A3) at (0,4) {\small $v^1_d$};
                \node[node] (A4) at (2,1) {};
                \node[node] (A5) at (2,3) {};
                
                \draw[entangled-strong] (A1) -- (A2);
                \draw[entangled] (A2) -- (A3);
                \draw[entangled] (A1) -- (A4);
                \draw[connection] (A2) -- (A4);
                \draw[entangled-strong] (A2) -- (A5);
                \draw[entangled] (A3) -- (A5);
                \draw[entangled] (A4) -- (A5);
                
                \node[gnn] (GNN1) at (0.3,0.3) {};
                \node[gnn] (GNN2) at (0.3,2.3) {};
                \node[gnn] (GNN3) at (0.3,4.3) {};
                \node[gnn] (GNN4) at (2.3,1.3) {};
                \node[gnn] (GNN5) at (2.3,3.3) {};
                
                \draw[path] (A1) -- (A2);
                
                \node[agent] (PA2) at (-0.5,2) {\small 1};

                \draw[message] (A1) to[bend right=20] node[message-label, pos=0.5] {} (A2);
                \draw[message] (A2) to[bend right=20] node[message-label, pos=0.5] {} (A1);
                \draw[message] (A2) to[bend right=15] node[message-label, pos=0.5] {} (A3);
                \draw[message] (A3) to[bend right=15] node[message-label, pos=0.5] {} (A2);
                \draw[message] (A1) to[bend right=15] node[message-label, pos=0.5] {} (A4);
                \draw[message] (A4) to[bend right=15] node[message-label, pos=0.5] {} (A1);
                \draw[message] (A2) to[bend right=15] node[message-label, pos=0.5] {} (A4);
                \draw[message] (A4) to[bend right=15] node[message-label, pos=0.5] {} (A2);
                \draw[message] (A2) to[bend right=20] node[message-label, pos=0.5] {} (A5);
                \draw[message] (A5) to[bend right=20] node[message-label, pos=0.5] {} (A2);
                \draw[message] (A4) to[bend right=15] node[message-label, pos=0.5] {} (A5);
                \draw[message] (A5) to[bend right=15] node[message-label, pos=0.5] {} (A4);
                \draw[message] (A3) to[bend right=15] node[message-label, pos=0.5] {} (A5);
                \draw[message] (A5) to[bend right=15] node[message-label, pos=0.5] {} (A3);
            \end{tikzpicture}
        }
    \end{minipage}
    \caption{Monitoring of the quantum topology. The green arrows depict the flow of monitoring information, which is either directly used by central path planning or aggregated using a \gls*{gnn}. The orange arrow depicts the control signal.}\label{fig:monitoring}
\end{figure}
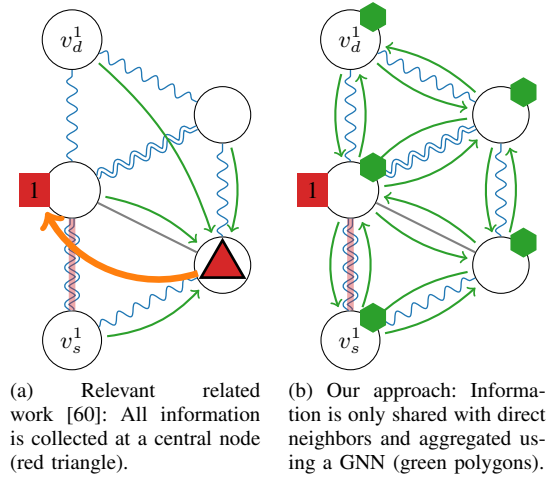
\endgroup
\subsection{Local monitoring information and message exchange}\label{subsec:node_observations}
While many quantum routing approaches from related work rely on some form of global information, this requires a central node to collect all information and coordinate the path planning, as shown in Fig.~\mbox{\ref{subfig:monitoring_centralized}}.
In contrast, we rely on local monitoring, as shown in Fig.~\mbox{\ref{subfig:monitoring_distributed}}.
In each step of the monitoring cycle (which can be decoupled from the steps of the environment), each node sends the message generated by its \gls*{gnn} to all of its one-hop neighbors in the physical topology.
In this work, we perform one monitoring cycle per step of the environment.
The input to each node consists of local monitoring information and the messages received from its direct neighbors.
It should be noted that nodes do not receive messages from distant nodes, but update their state from the messages received from their one-hop neighbors.
Due to the combination of all received messages in the new graph representation and the statefulness of the \gls*{gnn}, repeaters can obtain information about distant repeaters in the network over multiple monitoring cycles.
To allow for an association of received messages with each one-hop neighbor, they are added in the order in which the links appear in the action space of the agent (compare Sec.~\ref{subsec:agent}).

To increase the scalability of the approach, we utilize the idea of content-based addressing, in which each repeater knows if it is the target of a source-destination request.
This approach of content-based addressing relies heavily on the accuracy of the graph representations, as the target can otherwise only be found in the direct one-hop neighborhood.
Our local monitoring information is significantly different from that of Weil et al.~\cite{weil2024generalizability}, who utilize an encoding based on unique identifiers that is closely related to IP-based addressing.
Our change was necessary to apply the approach to real-world topologies: Both the varying node degree and the poor scaling of pure identifier-based addressing for learning-based approaches led to very poor performance in quantum networks, as showcased later in Sec.~\ref{subsec:training_process}.
Content-based addressing is very efficient for large networks.
It allows for better generalization to different graph sizes, as not the ID \textit{and} the location of a node in the graph, but only the location is considered.
For each node $v \in V$ and agent $i$, we provide target information, which is $1$ if $v^i_d = v$ (and $0$ otherwise).
If more than one source-destination pair is considered, then the observation differs for each source-destination pair, resulting in additional monitoring messages. Although this leads to additional network load, the network load of RELiQ is very low, as discussed in Sec.~\mbox{\ref{subsec:overhead}}.
Although content-based addressing scales worse with the number of source-destination pairs, we expect significantly fewer source-destination pairs than repeaters in a quantum network.

This information is augmented with the noise introduced by the quantum repeater and detailed information about the connected edges, which includes the number and maximum fidelity of the available elementary links and the target information of the connected nodes.
To address resource contention at the repeaters, the number of available elementary links considers the reservations of elementary links, which are created by other agents during path planning.

\subsection{Agent observations and actions}\label{subsec:agent}
The actions of an agent $i$ responsible for the source-destination pair $(v^i_s, v^i_d)$ depend on its current position $v^i_c$ in the graph, from which the agent can choose any edge connected to $v^i_c$.
Each agent attempts to establish one link between the source and destination at a time, but it is possible to assign multiple agents to the same source-destination pair to add multipath capabilities.
Note that elementary links are only reserved, but no swap operation is performed. The reservation is not specific to one specific elementary link, but indicates to other agents that fewer elementary links are available.
An action mask is used to prevent loops, i.e., all actions that would cause loops in the entanglement path are prohibited.
Although it is also possible for the model to learn this behavior, this can lead to suboptimal solutions and does not prevent the model from occasionally creating routing loops.
As proposed in \cite{weil2024generalizability}, we extend their approach to graphs with varying node degree using action masking~\cite{schneider2021distributed}: 
If the node degree of $v^i_c$ is below the maximum node degree supported by the model (i.e., the action space of the agent), the action mask is used to prevent the model from selecting those actions.

The agent observation consists of two parts: local monitoring information and the respective output of the \gls*{gnn} of the node $v^i_c$ where the agent is currently located.
Local monitoring information consists of a unique identifier of the associated source-destination pair, the age of the request in environment steps, and information about edges connected to its current location $v^i_c$.
For each possible edge, the swap probability of the connected quantum repeater, the number and maximum fidelity of the available elementary links, and the current status of the connected quantum repeater are provided.
The status indicates if the repeater has been visited before and if it is the source or destination of the source-destination pair $i$.
If the degree of the current node $v^i_c$ is below the maximum supported node degree, the observation is padded with zeros.

\subsection{Reward function and return}
For the reward function, we want to encourage two main properties of the routes selected by the agent: i) a high number of high-fidelity end-to-end entanglements, and ii) low resource consumption regarding elementary links.
Although fine-grained rewards are possible in this scenario, e.g., by considering the shortest available path in the quantum network, our early experiments showed that the benefits of these rewards are minimal and can even reduce the performance of the trained model.
Thus, we utilize sparse rewards, which are only non-zero if the agent is done.
This can happen for two reasons: i) the agent successfully found a path between the source and destination, or ii) the agent failed to find the path, that is, a maximum number of quantum repeaters $P_{max}$ was exceeded.
The reward function in Eq.~\ref{eq:reward} is used to incentivize agent $i$ to maximize the fidelity of the end-to-end entanglements.
It depends on the state of the environment $s_t$ at step $t$, the action the agent took $a_t^i$, and the next state of the environment $s_{t+1}$.

\begin{equation}\label{eq:reward}
    r^i(s_t, a_t^i, s_{t + 1}) = \begin{cases}
        F^i_{E2E} & \text{if success}\\
        F_{B} & \text{if path found}\\
        0 & \text{otherwise}\\
    \end{cases}
\end{equation}

The fidelity is maximized by directly considering the fidelity of the end-to-end entanglement $F^i_{E2E}$ in the reward.
If a path was found but no entanglement could be created, the agent is still provided with a reward equal to the baseline fidelity of $F_{B}=0.5$, which guides the agent toward the general goal during early training.
All intermediate rewards are $0$.

The reward is then used during training as part of the return. The return is the cumulative reward and considers not only current but also future rewards, allowing the agent to assess actions taken along the path.
Future rewards are discounted, encouraging shorter paths and lower consumption of elementary links.
We integrate multi-step learning into our approach, which has been shown to work well in previous works~\cite{10.5555/3504035.3504428}.
In our case, we choose to approximate the expected return via individual rollouts, as this has shown to drastically increase the convergence speed and performance of the trained models.
We calculate the return $R^i(\tau^i_{t:T})$ based on the states and actions $\tau^i_{t:T} = (s_t, a^i_t, s_{t+1}, a^i_{t+1}, \ldots, s_{T+1})$ taken by agent $i$ until the agent is reset after step $T$.
Eq.~\ref{eq:gamma} shows the definition of the return.

\begin{equation}\label{eq:gamma}
    R^i(\tau^i_{t:T}) = \sum_{k = t}^{T} \gamma^{k - t} \cdot r^i(s_k, a^i_k, s_{k+1})
\end{equation}

The discount factor $\gamma < 1$ is used to balance immediate and future rewards.
The exponential decay of future rewards incentivizes shorter paths and thus less resource consumption.

\section{Results}
\label{results}

\glsresetall




\begin{table}[tb]
    \centering
    \begin{tabular}{l r}
        \toprule[0.8pt]
        \textbf{Variable} & \textbf{Experiment Values} \\ \midrule[0.8pt]
        Number of repeaters & 10, 30, \textbf{100}, 300, 1000 \\ \hline
        Number of source-destination pairs & \textbf{1}, 3, 10, 30, 100 \\ \hline
        Attenuation constant $[\si{\decibel\per\kilo\meter}]$ & 0.15, \textbf{0.2}, 0.25 \\ \hline
        Initial fidelity & 0.8, 0.85, 0.9, \textbf{0.95}, 1.0 \\ \hline
        Average gate fidelity $f_\text{gate}$ & 0.9, 0.95, \textbf{1.0}\footnotemark \\ \hline
        Spread of $f_\text{gate}$ & 0.0, \textbf{0.1}, 0.2 \\ \hline
        Average number of decoupling pulses $n_\text{dec}$ & 64, 256, \textbf{1024} \\ \hline
        Spread of $n_\text{dec}$ & \textbf{0}, 256, 1024 \\ \hline
    \end{tabular}
    \caption{Parameters used in the evaluation. Default values are marked in bold.}
    \label{tab:eval_parameters_alt}
\end{table}
\footnotetext{Note that the gate fidelity of 50\% of the repeaters is below $1.0$ in the default setting, as the spread is set to $0.1$.}

\begin{figure*}[t]
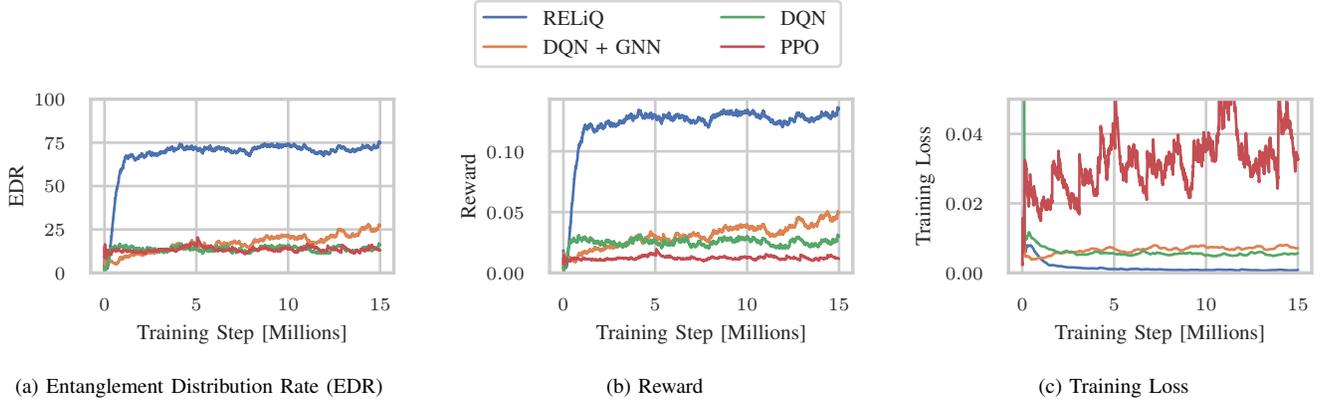

    \centering
    \input{figures/final_plots/legend_training.pgf}\\\vspace{-3mm}
    \begin{minipage}{\textwidth}
    \centering
    \subfloat[Entanglement Distribution Rate (EDR)]{
        \input{figures/final_plots/line___Train_Average_episode_packets.pgf}
    }\hfill
    \subfloat[Reward]{
        \input{figures/final_plots/line___Train_Reward.pgf}
    }\hfill
    \subfloat[Training Loss]{
        \input{figures/final_plots/line___Train_Loss.pgf}
    }
    \end{minipage}
    \caption{Training behavior of our approach compared to other reinforcement learning-based approaches.}
    \label{fig:training_visualization}
\end{figure*}

We evaluated our RELiQ algorithm against three learning-based approaches and six heuristic approaches for entanglement routing. For learning-based approaches, we compared RELiQ to a reinforcement learning-based approach with local view based on Deep Q-Learning~\cite{MKS+15} and Proximal Policy Optimization (PPO) with action masking support~\cite{Huang_2022}, as well as a reinforcement learning-based approach in conjunction with a \gls*{gnn}~\cite{weil2024generalizability}.
For heuristics, Q-PATH, Q-LEAP \cite{qpath_and_qleap}, GER, MGER, LBER, and NoNLBER \cite{vgraph2} were considered.
These algorithms have been chosen due to their varied approaches to routing: some rely on local information about available elementary links (GER, MGER, LBER, NoNLBER\footnote{NoNLBER utilizes all information of the two-hop neighborhood.}), while others leverage global information about available elementary links (Q-LEAP, Q-PATH).
For Q-LEAP and Q-PATH, we utilize a fidelity threshold of $0.7$, similar to the evaluation performed in \mbox{\cite{qpath_and_qleap}}.
Higher values lead to an increased fidelity, while lower values increase the number of created end-to-end entanglements slightly.
Decreasing this threshold below $0.55$ leads to a significantly reduced number of created end-to-end entanglements, caused by the gap between the predicted and actual fidelity of a created end-to-end entanglement, which leads to failed swap operations.
For the utility function used in Q-PATH and Q-LEAP for multiple source-destination pairs, we use $\alpha^*=\beta^*=0.5$ to balance resource consumption and degree of freedom as suggested by the authors.
For all approaches relying on quantum topology information for distant repeaters and edges, we implemented a monitoring system:
For Q-PATH and Q-LEAP, this monitoring system is based on messages to collect all information at a central coordination node, as shown in Fig.~\mbox{\ref{subfig:monitoring_centralized}}.
This induces a delay that depends on the distance to the coordination node.
For NoNLBER, we rely on a similar monitoring system to collect information about the two-hop neighborhood of all nodes, which is conceptually similar to the one-hop message exchange utilized by RELiQ.
RELiQ relies solely on local information and message passing, without prior information on the network topology or the availability of elementary links.

In our experiments, we analyzed the influence of different parameters on the algorithms as shown in Table \ref{tab:eval_parameters_alt}.
Note that we utilized a clipped Gaussian distribution for the gate fidelity of the individual repeaters.
Thus, the average gate fidelity is below $1$ for all scenarios in which the standard deviation of this fidelity is above $0$.
An analysis of the influence of the gate fidelity can be found in Appendix~\ref{subsec:influence_dep_error}.

The evaluation is performed for \SI{100}{} episodes for \SI{1000}{} steps each, with each step corresponding to \SI{10}{\milli\second} of simulation time.
We also experimented with other step durations, but found that the results follow similar patterns as with a step duration of \SI{10}{\milli\second}.
In each episode, a random graph structure is generated or a real-world topology is used, which has not been part of the training set for our learning-based approach and remains the same during one episode.
For each link in the network, elementary links are generated by simulating the physical properties of the links, as described in Sec.~\ref{qnetmodel}.
As the approaches create end-to-end entanglements, these elementary links are used, which might lead to physical connections without any elementary links available, effectively changing the quantum topology.
As the node degree of any node in the network may differ, we assume that the size of the quantum storage on each node increases linearly with the node degree.

The quality of the entanglements will degrade over time, which we modeled as described in Sec.~\mbox{\ref{qnetmodel}}.
In each step, each node can exchange messages with its one-hop neighbors.
Similarly, the path-finding for entanglement routing is performed at a rate of one hop per environment step.
However, swap operations along the path are assumed to be instantaneous to avoid unnecessary decay of the established end-to-end entanglements.
We use the Entanglement Distribution Rate (EDR) as our main metric for the performance of the algorithm and provide details on the fidelity that was achieved for one scenario.
In addition, we analyze the computational complexity and monitoring overhead for the different approaches.

The source code of our implementation is available online\footnote{\url{https://github.com/meusert/RELiQ}}.

\subsection{Training process and comparison to other learning-based approaches}\label{subsec:training_process}

\begin{figure}[t]
    \centering
    \input{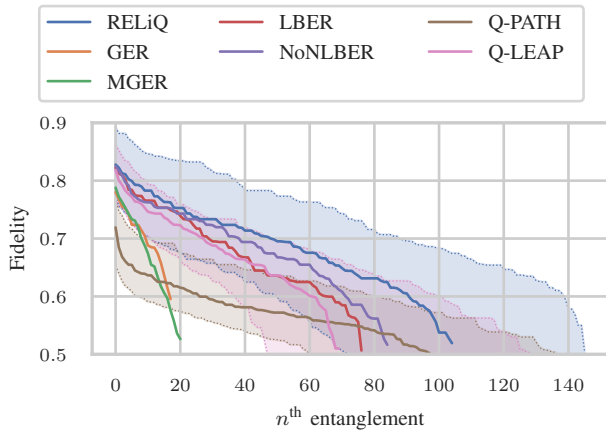}
    \caption{Median fidelity of the $n^\text{th}$ entanglement for the different approaches over 100 runs. The shadowed areas present the 25th and 75th percentile. For better readability, the percentiles are only plotted for RELiQ, Q-PATH, and Q-LEAP. If less than $n$ entanglements are created, the respective line ends early.}
    \label{fig:behavior_fidelity}
\end{figure}

In Fig.~\mbox{\ref{fig:training_visualization}}, we visualize the training for RELiQ, a DQN-based approach~\mbox{\cite{MKS+15}}, and DQN combined with GNN~\mbox{\cite{weil2024generalizability}}. In addition, we compare with a PPO extension that supports action masking~\mbox{\cite{Huang_2022}}, which is necessary to mask non-available links in the network.
All approaches were trained for \mbox{\SI{15000000}{}} steps and each training episode only lasted for \mbox{\SI{500}{}} steps to increase the number of unique topologies seen during training.
During training, we relied on an exponential decay of qubits in quantum memory, as this increases the unique fidelity values seen during training.
During evaluation, we rely on the realistic decay presented in Sec.~\mbox{\ref{qnetmodel}}.
For DQN-based approaches, we used $\epsilon = 1$ at the beginning and $\epsilon_\text{decay} = 0.99999$, a discount factor $\gamma = 0.95$, and a learning rate $\text{lr}=0.0005$.
We used a mini-batch size of $32$ and performed $200$ steps of the environment between training iterations.
An analysis of the influence of the different hyperparameters is performed in the Appendix~\mbox{\ref{subsec:hyperparams}}.
For PPO, an implementation of SB3 Contrib was used with the default parameters.
\parskip 0pt

Our results indicate that PPO, DQN, and DQN + GNN fail to converge on effective solutions: For the DQN and PPO agents, the randomness of the underlying network topologies prevents any improvements.
In most cases, the performance of PPO is similar to the performance of the DQN approach, which is caused by the limited view of these agents on the network.
This is an inherent issue with decentralized approaches, which is independent of the utilized reward function.

For DQN + GNN, the structure of the observation used prevents convergence, as each node is addressed by its unique identifier using a one-hop encoding, leading to huge observations that prevent generalization.
Compared to the observation space of RELiQ, which is of size $49$ for each agent and $34$ for each node with $100$ nodes, the DQN + GNN approach with addressing implemented in~\cite{weil2024generalizability} uses an observation space of size $908$ for each agent and $705$ for each node.

The training process of RELiQ is relatively stable, with a clear improvement in the first \SI{2000000}{} training steps.
After that, the EDR and reward improve only slightly (from an average EDR of $66.8$ at \SI{2000000}{} steps to $74.9$ at \SI{15000000}{} steps), but the loss still decreases, as the model obtains a better understanding of the underlying graph structures.

\subsection{Comparison with baseline approaches}
In this section, we compare the performance of RELiQ with the six heuristics. The other learning-based approaches are not included because of their low performance.
It should be noted that all experiments except those that compared performance for multiple source-destination pairs were performed with the same model, highlighting the high generalization capabilities of RELiQ.
For multiple source-destination pairs, the training was performed in a multi-agent environment, in which the same model controlled all agents simultaneously to learn coordination between the agents.
For the experiment with multiple source-destination pairs, a single model was used for all settings, which was trained with $10$ concurrent agents.

\begin{figure*}[t]
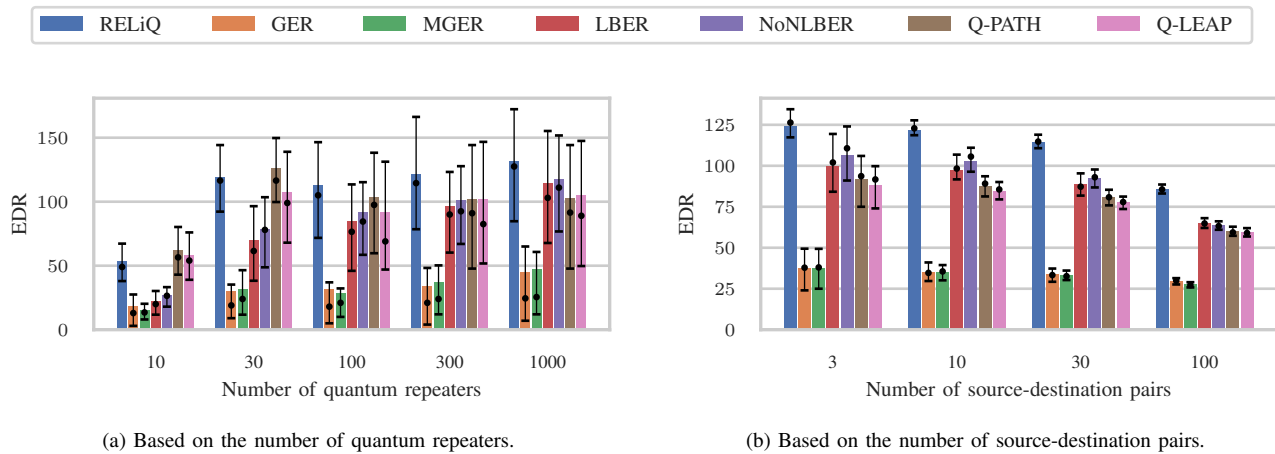

    \centering
    \input{figures/final_plots/legend.pgf}
    \subfloat[Based on the number of quantum repeaters.\label{fig:packets_per_network_size}]{\input{figures/final_plots/bar___n_router_connected___average_episode_packets___no_legend.pgf}}\quad
    \subfloat[Based on the number of source-destination pairs.\label{fig:packets_per_source_destination_pairs}]{\input{figures/final_plots/bar___n_data___average_episode_packets___no_legend.pgf}}
    \caption{Entanglement Distribution Rate (EDR) based on the scale of the quantum network.}
\end{figure*}

The fidelity of established end-to-end entanglements is essential for the usability of the respective entanglement.
Thus, Fig.~\ref{fig:behavior_fidelity} displays the fidelity of the $n^\text{th}$ highest entanglement for all approaches over \SI{100}{} runs.
If an approach creates less than $n$ entanglements in any run, no value is considered for that run.
The results clearly show that the approaches can be grouped into approximately three groups: GER and MGER struggle to create many entanglements, leading to a rapid drop in fidelity. 
Interestingly, Q-PATH and Q-LEAP also struggle to create high-fidelity entanglements, even though they are configured to guarantee a fidelity of $0.7$.
This is caused by stale information about the quantum topology and the heterogeneity of quantum repeaters. The latter is especially impactful, as Q-PATH and Q-LEAP both use multiplication for an estimate of the swap output fidelity, which is not accurate for low and diverse gate fidelities of quantum repeaters. LBER and NoNLBER perform significantly better in most cases because of their iterative path planning and fast reactivity to changes in the quantum topology. RELiQ outperforms all other approaches significantly, especially for high values of $n$.

We further analyzed the scalability of these algorithms based on network size, that is, the number of repeaters.
As shown in Fig.~\ref{fig:packets_per_network_size}, RELiQ significantly outperforms all tested heuristics in terms of the entanglement distribution rate (EDR), including heuristics requiring global information about the network and non-local updates.
In each bar plot in this section, the bar highlights the average value achieved over \SI{100}{} runs, while the error bar visualizes the median and the $75^\text{th}$ and $25^\text{th}$ percentiles, respectively.
With the exception of Q-PATH and Q-LEAP, the performance of all approaches increases with increasing network size, which can be explained by the higher number of backup paths available for larger networks.
For Q-PATH and Q-LEAP, the increasing age of obtained information about the quantum topology heavily impacts their performance.
The performance increase for RELiQ is especially promising, as the improved scalability of decentralized approaches often reduces their performance.
Through our observation modeling explained in Sec.~\ref{subsec:node_observations}, we avoid this issue and allow similar performance in all settings.
This is crucial for realistic settings where scaling can pose a significant challenge for efficient routing implementations.

The content-based addressing scheme employed by RELiQ scales worse with the number of source-destination pairs compared to identifier-based approaches.
The computational load and monitoring overhead of RELiQ are discussed in Sec.~\mbox{\ref{subsec:overhead}}.
Fig.~\mbox{\ref{fig:packets_per_source_destination_pairs}} shows the performance of the different approaches for different numbers of source-destination pairs.
To allow the underlying topology to handle up to $100$ parallel source-destination pairs and the correlated demand of elementary links, we perform this analysis in a network with $1000$ quantum repeaters.
It can be observed that the performance of all approaches decreases with an increasing number of source-destination pairs due to the increasing resource contention.
The number of created elementary links per source-destination pair initially decreases slightly with increasing source-destination pairs, but later decreases significantly due to resource contention.
For all cases, the performance of RELiQ remains above the performance of the baseline approaches, including Q-PATH and Q-LEAP.

\begin{figure*}[t]
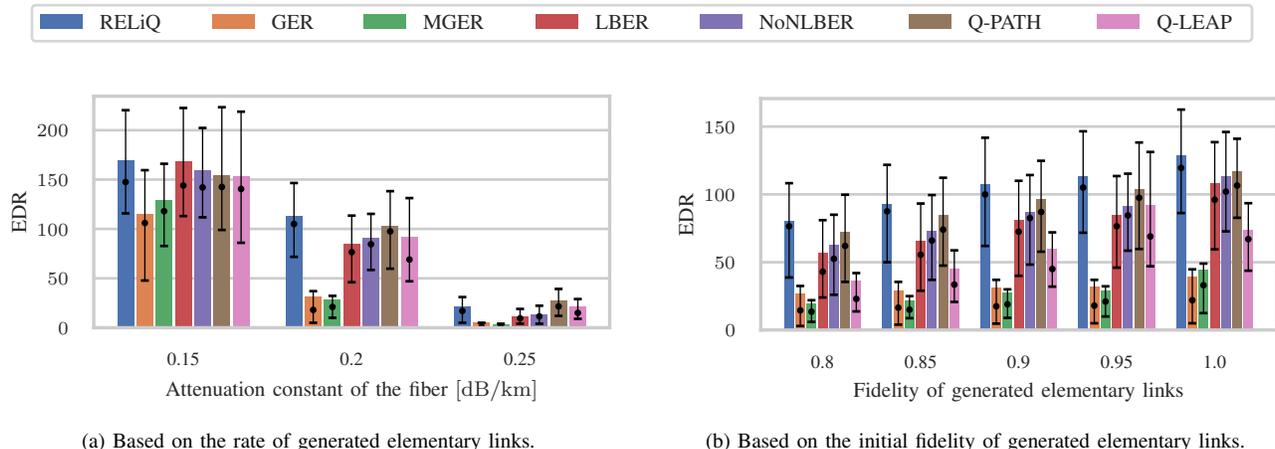

    \centering
    \input{figures/final_plots/legend.pgf}
    \subfloat[Based on the rate of generated elementary links.\label{fig:packets_per_link_prob}]{\input{figures/final_plots/bar___attenuation_coefficient___average_episode_packets___no_legend.pgf}}\quad
    \subfloat[Based on the initial fidelity of generated elementary links.\label{fig:packets_per_initial_fidelity}]{\input{figures/final_plots/bar___initial_fidelity___average_episode_packets___no_legend.pgf}}
    \caption{Entanglement Distribution Rate (EDR) based on the generation of elementary links.}
\end{figure*}

While the creation of entangled links between neighboring repeaters is probabilistic, the success probability is affected by the fiber quality, which we represent through the attenuation constant.
If the attenuation constant is low, more entanglements can be created, i.e., the approaches are less limited by the availability of elementary links.
We observed (Fig.~\ref{fig:packets_per_link_prob}) that in the three tested scenarios, RELiQ either surpassed or equaled the two global information heuristics and the local information ones.
Interestingly, NoNLBER performs very well at low attenuation, suggesting that the per-hop path-finding of NoNLBER can outperform global-information approaches if the availability of elementary links is sufficient.

In addition, the fidelity of generated entanglements may vary depending on the specific generation methodology used.
In Fig.~\mbox{\ref{fig:packets_per_initial_fidelity}}, we analyze the influence of this fidelity on the routing of entanglements.
It should be noted that fidelities below $0.95$ are generally very limited for quantum routing, as the maximum number of swap operations is relatively low (compare Fig.~\mbox{\ref{fig:dep_error_routing}}).
To address this, we performed entanglement distillation similarly to Q-PATH and Q-LEAP, but only once the entanglements are created, to increase the fidelity whenever possible.
We distill elementary links up to a fidelity of $0.98$, which was found empirically by considering the performance of the local knowledge-based approaches.
This leads to similar results as an increased attenuation coefficient, as fewer high-fidelity qubits are available for routing.
For Q-PATH and Q-LEAP, this mechanism is disabled to avoid interfering with their built-in distillation protocol.
However, due to the staleness of obtained information in Q-PATH and Q-LEAP, the distillation decisions are often not optimal, leading to failed swap operations or excessive consumption of elementary links.
Interestingly, Q-LEAP performs poorly for an initial fidelity of $1$. This is caused by the utilization of the product of fidelities along the path, resulting in very long paths and many swap operations.
For all fidelities, RELiQ outperforms all other approaches significantly, including Q-LEAP and Q-PATH, which follow their own distillation protocol.

\begin{figure*}
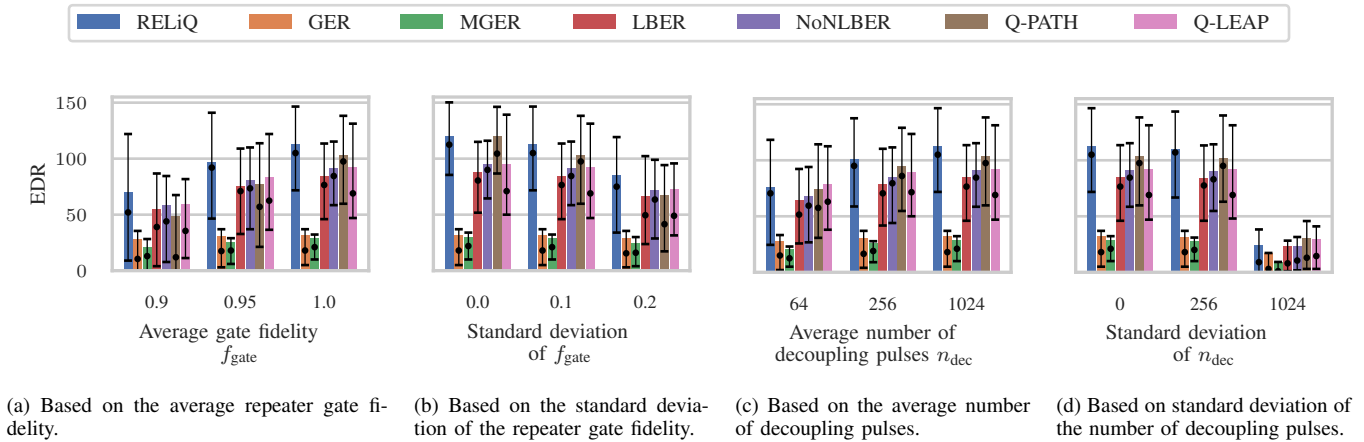

    \centering

    \input{figures/final_plots/legend.pgf}
    
    \subfloat[Based on the average repeater gate fidelity.\label{fig:packets_per_avg_swap}]{\input{figures/final_plots/bar___swap_prob___average_episode_packets___no_legend.pgf}}\quad
    \subfloat[Based on the standard deviation of the repeater gate fidelity.\label{fig:entanglement_prob_average_episode_packets}]{\input{figures/final_plots/bar___swap_prob_std___average_episode_packets___no_legend.pgf}}\quad
    \subfloat[Based on the average number of decoupling pulses.\label{fig:packets_per_decoupling_pulses_avg}]{\input{figures/final_plots/bar___decoupling_pulses_avg___average_episode_packets___no_legend.pgf}}\quad
    \subfloat[Based on standard deviation of the number of decoupling pulses. \label{fig:packets_per_decoupling_pulses_std}]{\input{figures/final_plots/bar___decoupling_pulses_std___average_episode_packets___no_legend.pgf}}
    
    \caption{Entanglement Distribution Rate (EDR) per episode as a function of the quality of the quantum repeaters.}
\end{figure*}

Given the high dependence of quantum networking on the quality of the quantum repeaters, we assessed the impact of different gate fidelities (for details, see Appendix~\ref{subsec:influence_dep_error}).
The results (Fig. \ref{fig:packets_per_avg_swap}) indicate that RELiQ outperforms the tested heuristics for a high average gate fidelity, while a lower gate fidelity appears to reduce the performance gap.
Even a slightly reduced gate fidelity drastically reduces the maximum number of swap operations that produce usable entanglements, making it impossible to establish end-to-end entanglements over many repeaters and preventing (often longer) backup paths.
The performance of RELiQ decreases significantly for $f_\text{gate}=0.9$, but all heuristics perform poorly under these conditions.
\parskip 0pt

To further evaluate the robustness of the approaches, we investigated the effect of a high standard deviation of the gate fidelity.
This scenario represents the realistic case where not all repeaters have the same quality or operate at optimal parameters.
In this context (see Fig.~\ref{fig:entanglement_prob_average_episode_packets}), RELiQ handles heterogeneity in the gate fidelities very well, while Q-PATH and Q-LEAP both struggle with increasing heterogeneity.


Further analysis considered the impact of qubit decay rates in quantum memories, which can negatively affect quantum network functionality. As elementary links are formed between neighboring nodes, the shared entangled qubits are stored in quantum memories. The decay of qubits in quantum memories can be influenced by decoupling pulses as described in Sec.~\mbox{\ref{qnetmodel}}.
Fig.~\mbox{\ref{fig:packets_per_decoupling_pulses_avg}} shows the influence of the number of decoupling pulses on the performance of the approaches.
For a low number of decoupling pulses, all approaches struggle to create end-to-end entanglements due to the fast decoherence of the stored qubits (compare Fig.~\mbox{\ref{fig:decay}}).
However, for higher values, the performance of all approaches increases, with RELiQ outperforming the other approaches for values above $64$.
For $n_\text{dec}=64$, the qubits decay rapidly and all approaches rely on freshly generated entanglements during path planning, drastically limiting their performance.
We expect that distillation methods can further improve performance in such situations, especially when combined with local decision making.

Fig.~\mbox{\ref{fig:packets_per_decoupling_pulses_std}} shows the influence on the heterogeneity of quantum memories in the repeaters.
Interestingly, the influence of changes to the standard deviation is quite minor up to a standard deviation of $256$ and the approaches remain largely unaffected.
However, for a standard deviation of $1024$, the performance of all approaches decreases significantly, with Q-PATH and Q-LEAP performing the best and RELiQ performing slightly worse.

\begin{table*}[t]
\centering
\begin{tabular}{c | c | c c c c c c c c c c || c c c c}
\toprule[0.8pt]
\multicolumn{2}{c|}{} & \multicolumn{10}{c||}{ Local Knowledge } & \multicolumn{4}{c}{Global Knowledge} \\ 
\multicolumn{2}{c|}{}  & \multicolumn{2}{c}{RELiQ} & \multicolumn{2}{c}{GER} & \multicolumn{2}{c}{MGER} & \multicolumn{2}{c}{LBER} & \multicolumn{2}{c||}{NoNLBER} & \multicolumn{2}{c}{Q-PATH} & \multicolumn{2}{c}{Q-LEAP}\\
\multicolumn{2}{c|}{} & \textit{EDR} & \textit{Fid.}& \textit{EDR} & \textit{Fid.}& \textit{EDR} & \textit{Fid.}& \textit{EDR} & \textit{Fid.}& \textit{EDR} & \textit{Fid.}& \textit{EDR} & \textit{Fid.}& \textit{EDR} & \textit{Fid.}\\ \midrule[0.8pt]
\multirow{4}{*}{\rotatebox{90}{\parbox{1cm}{ SNDlib \cite{sdnlib}}}} & Cost 266 & \textbf{137.3} & \textbf{79}\% & 79.6 & 70\% & 73.9 & 68\% & 115.6 & 76\% & 122.5 & 78\% & 131.2 & 65\% & 124.6 & 74\%\\ \cline{2-16}
& Germany & \textbf{175.5} & \textbf{82}\% & 133.6 & 78\% & 138.3 & 76\% & 167.6 & 80\% & 168.4 & 81\% & 169.1 & 67\% & 170.9 & 76\%\\ \cline{2-16}
& EU & \textbf{152.2} & \textbf{82}\% & 94.9 & 72\% & 88.8 & 73\% & 139.8 & 80\% & 137.6 & 81\% & 146.5 & 66\% & 142.2 & 76\%\\ \cline{2-16}
& Poland & 153.5 & \textbf{85}\% & 63.7 & 75\% & 66.3 & 70\% & 123.8 & 83\% & 128.1 & 84\% & \textbf{155.0} & 72\% & 151.4 & 74\%\\ \hline
\multirow{6}{*}{\rotatebox{90}{\parbox{1.8cm}{Internet Topology Zoo \cite{topozoo}}}} & US & \textbf{139.6} & \textbf{80}\% & 96.8 & 78\% & 97.5 & 76\% & 115.7 & 76\% & 127.5 & 78\% & 119.6 & 64\% & 119.7 & 73\%\\ \cline{2-16}
& Finland & 54.7 & \textbf{73}\% & 41.0 & 70\% & 39.1 & 60\% & 44.0 & 68\% & 43.9 & 61\% & \textbf{61.3} & 61\% & 50.9 & 68\%\\ \cline{2-16}
& Poland & 114.6 & \textbf{79}\% & 99.8 & 77\% & 99.4 & 77\% & 102.4 & 77\% & 106.8 & 76\% & \textbf{121.1} & 63\% & 111.3 & 74\%\\ \cline{2-16}
& UK & 228.9 & \textbf{87}\% & 227.6 & 85\% & 227.5 & 85\% & 227.5 & 85\% & 228.0 & 85\% & \textbf{231.9} & 72\% & 229.1 & 74\%\\ \cline{2-16}
& Canada & \textbf{163.6} & \textbf{81}\% & 129.1 & 77\% & 124.6 & 78\% & 154.8 & 79\% & 156.1 & 80\% & 157.7 & 66\% & 156.9 & 76\%\\ \cline{2-16}
& York & 153.0 & \textbf{79}\% & 148.8 & 77\% & 148.7 & 77\% & 148.6 & 78\% & 149.3 & 78\% & 145.2 & 65\% & \textbf{154.2} & 74\%\\ \hline
\end{tabular}

\caption{Comparative evaluation of RELiQ and the considered heuristic algorithms on a series of real network topologies. For every algorithm the average number of successfully routed packets (end-to-end entangled pairs) is presented, as well as the average fidelity of the final states. 
The highest values for EDR and fidelity are marked bold.
}
\label{table:benchmarks}
\end{table*}

We also evaluated RELiQ on several real-world network topologies reflecting potential quantum network implementations. We retrieved and managed those topologies using Topohub~\cite{Topohub} and added additional quantum repeaters to prevent excessively long links. The results, presented in Table \ref{table:benchmarks}, are divided between local and global information algorithms. In most scenarios, RELiQ outperforms all baseline heuristics regarding the EDR or the achieved fidelity, often outperforming all other approaches in both categories. 
As the length of links influences the number of elementary links, the performance can differ significantly, as also shown in Fig.~\ref{fig:packets_per_link_prob}. NoNLBER performs very well in some topologies like UK or York, which is justified by the high availability of elementary links in these topologies due to their comparably short links.
Similarly, Q-PATH and Q-LEAP outperform RELiQ in some topologies with a low number of quantum repeaters.
As already shown in the analysis for the network size in Fig.~\mbox{\ref{fig:packets_per_network_size}}, Q-LEAP and Q-PATH perform very well if the network is small.
However, even in these topologies, RELiQ consistently achieved higher fidelity compared to both Q-LEAP and Q-PATH.

Overall, the robustness of RELiQ to the varying gate fidelities of quantum repeaters, attenuation constants, qubit memory lifetimes, and network structures, coupled with its design of using only local information without any global network information, underscores the strength of our proposed solution for realistic quantum network implementations.

\subsection{Computational complexity and monitoring overhead}\label{subsec:overhead}

\begin{figure}[t]
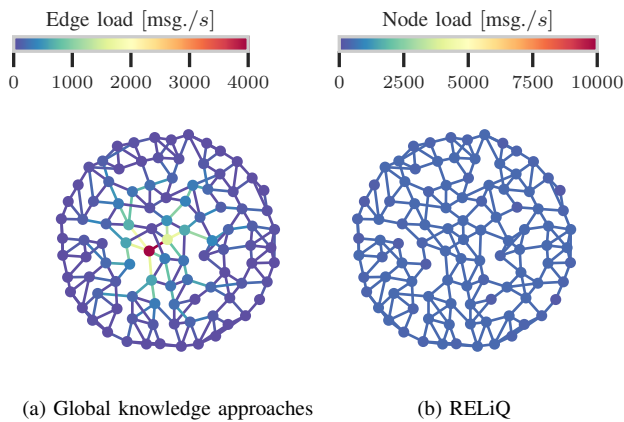

    \centering
    \input{figures/final_plots/topology_100___legend_edges.pgf}\quad
    \input{figures/final_plots/topology_100___legend_nodes.pgf}\\[-.6cm]
    \subfloat[Global knowledge approaches]{\input{figures/final_plots/topology_runtime-0--router-100--qpath.pgf}}
    \subfloat[RELiQ]{\input{figures/final_plots/topology_runtime-0--router-100.pgf}}\quad
    \caption{Message load per individual repeater in a network of 100 quantum repeaters considering both forwarding and processing of messages. While the overall number of messages is comparable between the scenarios, RELiQ distributes the load evenly between all repeaters and links in the network.}
    \label{fig:load_topology}
\end{figure}

\begin{figure}
    \centering
    \input{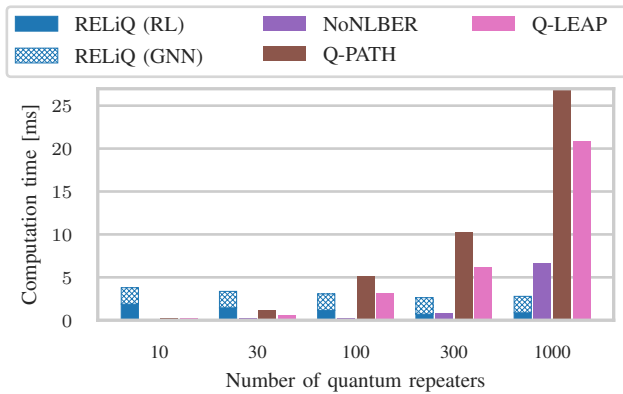}
    \caption{Runtime of the different approaches for each step of simulation.}
    \label{fig:computational_complexity}
\end{figure}

RELiQ relies on iterative message exchange between one-hop neighbors to generate global graph representations.
Thus, messages are not forwarded further than the one-hop neighbors, heavily limiting the number of messages received and sent by each repeater: Each repeater receives one message per source-destination pair from each of its neighbors.
In this work, we used messages consisting of $32$ floats, resulting in $256$ byte messages with each float being encoded through $8$ byte.
These messages are transmitted by all nodes to all of their one-hop neighbors in each step of the environment, resulting in two messages per link and source-destination pair.
With a step duration of \mbox{\SI{10}{\milli\second}}, the load on each link is as low as $409.6$ kbps for each source-destination pair without any additional optimizations, such as compression.
Reducing the message size by $50\%$ reduces the performance by approx. $4.5\%$.
While it is also possible to use RELiQ with monitoring messages as small as $8$ byte, this leads to a significant performance reduction of approx. $17\%$.
Compared to global knowledge approaches, the load is much more evenly distributed among repeaters and links, as also shown in Fig.~\mbox{\ref{fig:load_topology}}.
Although this leads to a lower load on each individual repeater, it also has advantages regarding failure tolerance and the actuality of information on the quantum topology.
Even though the number of source-destination pairs increases the total number of messages processed by each repeater, the number and size of received messages remains negligible compared to the fiber bandwidth of $\geq 1$ Gbps.
Similarly, the computational load of RELiQ is relatively low as shown in Fig.~\mbox{\ref{fig:computational_complexity}}.
For RELiQ, we divide the computation time between the reinforcement learning and the \mbox{\gls*{gnn}}.
While the computation time of all approaches increases significantly with network size, the distributed nature of RELiQ leads to a constant load on each repeater, which is independent of network size.
For small networks, the relatively constant inference time of RELiQ is much higher compared to the baseline approaches.
RELiQ outperforms Q-PATH and Q-LEAP beyond $100$ repeaters, and NoNLBER at $1000$ repeaters.
Interestingly, the computation time of RELiQ even decreases slightly, which is likely due to the longer paths found in larger networks and the decreased number of agent resets.

\section{Discussion}
\label{discussion}

\glsresetall

In this section, we discuss the main properties of our proposed RELiQ approach and highlight its benefits and limitations.
Although learning-based algorithms generally struggle with rapidly changing topologies, RELiQ can handle the dynamic nature of quantum networks, which are characterized by the depletion of elementary links.
Thus, the graph structure is constantly changing, and simple algorithms that mainly consider the topology of the underlying network, such as GER and MGER, achieve a relatively low number of end-to-end entanglement pairs.
The results presented in the previous section clearly show that our approach performs better than algorithms that utilize the full information of the available elementary links in most scenarios.
This is because these approaches need to rely on potentially outdated information, whereas RELiQ learns to encode the quantum network state and respond to it accordingly.
By defining the reward function in RELiQ based on the quality of end-to-end entanglements, our approach learns to find efficient paths in the network and to exchange the necessary information between repeaters.

RELiQ performs well on previously unseen topologies, as can be seen throughout the evaluation. 
In addition, RELiQ also performs well on real communication network topologies, as shown in Table~\ref{table:benchmarks}, further highlighting its adaptability to unseen networks.
Note that the performance of RELiQ with respect to EDR is sometimes slightly worse compared to either NoNLBER, Q-PATH, or Q-LEAP on some topologies.
However, even in these situations, RELiQ requires no retraining and performs well; its EDR remains competitive within a few percent of the top-performing approaches, while it consistently surpasses all others in the fidelity of the generated entanglements.

\section{Conclusion}
\label{conclusion}
\glsresetall
In this paper, we introduced RELiQ, a novel quantum routing algorithm based on reinforcement learning, designed to enhance the robustness and efficiency of entanglement distribution in quantum networks.
Our approach addresses the inherent challenges posed by quantum network operations, such as link decay, fidelity loss, and the probabilistic nature of quantum processes, by utilizing a learning mechanism that is trained centrally but relies solely on local information exchange during inference.

We structured our method around a recurrent message-passing framework that enables nodes to iteratively exchange and aggregate local observations to build comprehensive global representations of the network. This framework supports decision-making processes in the absence of global topology information, which is a significant improvement over many existing heuristic approaches that require extensive global information.
This design choice also enables the algorithm to adapt to various physical topologies and dynamic quantum network conditions without requiring retraining.

The performance of RELiQ was rigorously evaluated against six well-established heuristic algorithms, including local and global information-based methods, in both random and real-world topologies.
RELiQ demonstrated high scalability by efficiently routing entangled pairs across networks of varying sizes. It consistently exhibited higher resilience under varying conditions of entanglement swaps, elementary link establishment, and quantum memories, maintaining high performance with respect to the number of successfully routed entangled pairs and the overall fidelity of the entangled states.
In real-world network topologies, RELiQ outperformed all local-information heuristics in terms of the average number of packets routed per episode and maintained high fidelity of the final entangled states, comparable to or better than global-information heuristics, showcasing its practical applicability in realistic quantum network scenarios.
In future work, we plan to extend our approach to multipartite entanglements and explicit distillation decisions.

\bibliographystyle{IEEEtran}
\bibliography{bibliography2}

\appendix 
\setcounter{equation}{0} 
\renewcommand{\theequation}{A\arabic{equation}} 

\section{Appendix}
\label{appendix}

\subsection{Depolarization error}\label{appendix_error}
Various noise models prevalent in quantum photonic devices have previously been studied~\cite{Vischi_2024} \cite{king2002capacityquantumdepolarizingchannel}. 
In Fig.~\ref{fig:Qiskit_Circuit}, we specifically inject noise after the creation of Bell pairs using a two-qubit depolarization channel~\cite{PhysRevA.60.1888}.
The depolarization channel $\mathcal{E}$ with a depolarization error $\lambda$ is defined as: 

\begin{equation} \label{Depolarize}
    \mathcal{E}(\rho) = (1-\lambda)\rho + \lambda \mathrm{Tr} [\rho]\frac{I}{2^m}
\end{equation}
where $\text{Tr}[\rho]$ is the trace of the density matrix, $I$ the identity matrix, and $m$ the number of qubits.

\begin{figure}[b]
    \centering
    \input{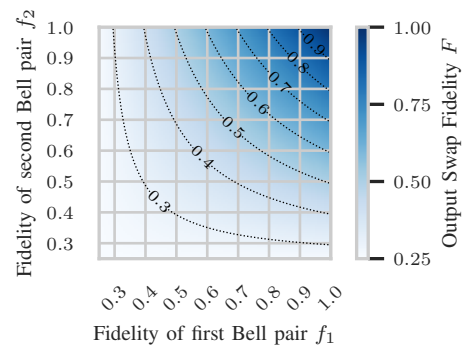}
    \caption{Output swap fidelity $F$ in terms of the input fidelity values $f_1$ and $f_2$ with $f_\text{gate}=1$.}
    \label{fig:HeatMap}
\end{figure}

\begin{figure*}[htb]
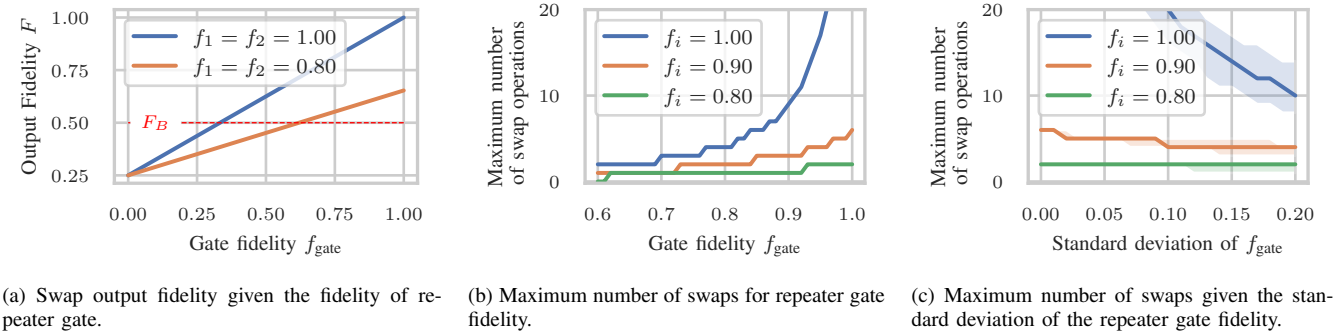

    \centering
    \subfloat[Swap output fidelity given the fidelity of repeater gate.\label{fig:Dep_var}]{\input{figures/final_plots/fidelity_emulation.pgf}}\quad
    \subfloat[Maximum number of swaps for repeater gate fidelity.]{\input{figures/final_plots/line___max_path_length___fidelities.pgf}\label{fig:max_number_of_hops}}\quad
    \subfloat[Maximum number of swaps given the standard deviation of the repeater gate fidelity.]{\input{figures/final_plots/line___max_path_length_std___fidelities.pgf}\label{fig:max_number_of_hops_std}}
    \caption{Influence of the fidelity of repeater gate on entanglement routing.\label{fig:dep_error_routing}}
\end{figure*}

The expression of fidelity for a noisy state $\mathcal{E}(\rho)$ and an ideal state $\rho$ can then be written as: $f (\mathcal{E}(\rho))= \mathrm{Tr}(\rho \mathcal{E} (\rho))$ \mbox{\cite{NIELSEN2002249}}. The output states from these two noisy error channels act as input states for the repeater. We establish a relation between the fidelity and depolarization error in order to get these noisy states that can be fed into the repeater.

From Eq.~\ref{Depolarize}, substituting $ \mathcal{E}(\rho) $ into the fidelity formula yields the fidelity from each error channel:

$$
f = \text{Tr}\left(\rho \left[(1 - \lambda)\rho + \lambda \, \mathrm{Tr}[\rho] \frac{I}{2^m} \right] \right).
$$

Using the properties of the trace:
$$
f = \text{Tr}((1 - \lambda)\rho^2) + \text{Tr}\left(\lambda\rho \frac{I}{2^m}\right).
$$

$$
f = (1 - \lambda) + \lambda \, \text{Tr}\left(\rho \frac{I}{2^m}\right).
$$

Since $ \displaystyle \text{Tr}\left(\rho \frac{I}{2^m} \right ) = \frac{1}{2^m} \text{Tr}(\rho I) = \frac{1}{2^m}$, for $ m = 2 $ (two-qubit gate):

$$
f = (1 - \lambda) + \frac{\lambda}{4} \implies \lambda =\frac{4}{3}(1-f). 
$$
 
Thus, the two input Bell pairs are prepared and coded in qiskit using this two-qubit depolarization error $\lambda$ with varied fidelities $f_1$ and $f_2$ for the first and second depolarization error channels $\lambda_1$ and $\lambda_2$, respectively. The repeater performs the Bell state measurements and outputs the final noisy state that creates an entanglement between nodes A and B. The final reduced density matrix $\rho_f$ for this entangled state $\rho_{AB}$ can be calculated using \mbox{Eq.~\ref{Depolarize}}: 
\begin{equation} \label{Depolarize2}
    \rho_{f} =  \rho_{AB} (f_\text{gate}) + (1 - f_\text{gate}) \frac{I_{AB}}{4}
\end{equation}
where $f_\text{gate}$  represents the  fidelity of the repeater section. The $\rho_{f}$ can be compared to the ideal density matrix of the Bell state $(\sigma_{AB})$ to calculate the final output fidelity $F$ of the entanglement swapping circuit using the formula: $F(\rho_{f},\sigma_{AB}) = \left(\text{Tr}\left(\sqrt{\sqrt{\rho_{f}}\sigma_{AB} \sqrt{\rho_{f}}}\right)\right)^2$ \cite{Nielsen_Chuang_2010}.

\begin{figure*}[thb]
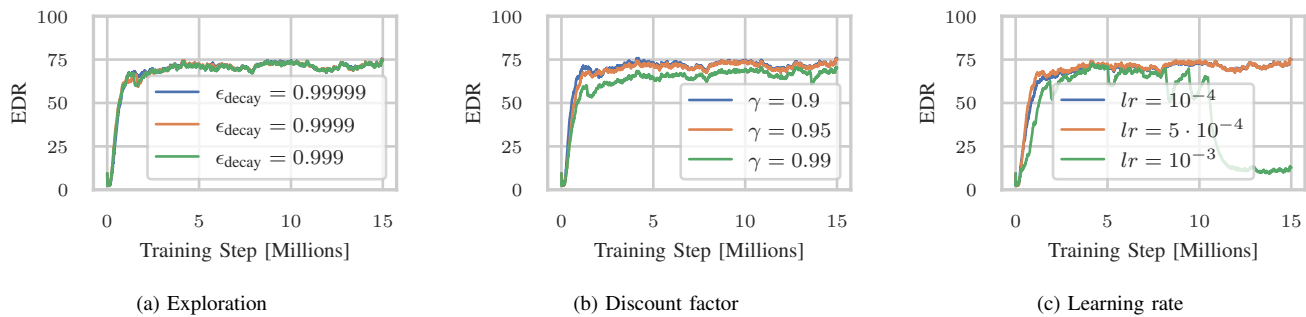

    \subfloat[Exploration\label{fig:exploration}]{\input{figures/final_plots/line___hyperparam_epsilon__Train_Average_episode_packets.pgf}}\quad
    \subfloat[Discount factor]{\input{figures/final_plots/line___hyperparam_gamma__Train_Average_episode_packets.pgf}}\quad
    \subfloat[Learning rate]{\input{figures/final_plots/line___hyperparam_lr__Train_Average_episode_packets.pgf}}
    \caption{Hyperparameters}\label{fig:hyperparams}
\end{figure*}



\subsection{Influence of the repeater gate fidelity $(f_\text{gate})$}\label{subsec:influence_dep_error}
Fig.~\ref{fig:HeatMap} shows the output swap fidelity $F$ for an ideal quantum repeater ($f_\text{gate}=1$) given the input fidelities $f_1$ and $f_2$ of the two Bell pairs. It is evident that the output fidelity $F$ is maximized when both input pairs possess high fidelity; furthermore, $F$ equals the fidelity of one input pair if the other has perfect fidelity ($f_1=1$ or $f_2=1$). However, even under these ideal repeater conditions ($f_\text{gate}=1$), $F$ decreases significantly as input fidelity degrades. For example, two Bell pairs with $f_1 = f_2 = 0.8$ yield $F \approx 0.65$, a substantially reduced output swap fidelity. If the two Bell pairs have $f_1=f_2=0.7$, the resulting fidelity $F\approx 0.52$ renders the entanglement almost unusable after only a single swap.

As shown in \mbox{Fig.~\ref{fig:dep_error_routing}}, the output fidelity of a swap operation decreases significantly with decreasing gate fidelity $f_\text{gate}$. Lower values of $f_\text{gate}$ often lead to an output fidelity below the threshold $F_{B}=0.5$, as evident in \mbox{Fig.~\ref{fig:Dep_var}}. If the fidelity of an entanglement drops below $F_{B}=0.5$, it cannot be distilled and is therefore unusable for quantum applications. For example, although the output swap fidelity for $f_1 = f_2 = 0.80$ is sufficiently high for an ideal repeater ($f_\text{gate}=1$), the output fidelity drops below $F_{B}$ once $f_\text{gate} \lessapprox 0.65$.

\balance

\mbox{Fig.~\ref{fig:max_number_of_hops}} further extends this analysis, illustrating the maximum number of sequential swaps possible along a path before the end-to-end fidelity drops below $0.5$. This analysis assumes that the fidelity $f_i$ of all initial links and the gate fidelity of all repeater gates along the path are equal. It can be seen that higher gate fidelity severely restricts the maximum possible path length for entanglement routing. This detrimental effect of the gate fidelity is especially impactful when combined with low initial elementary link fidelities, thereby highlighting the critical limitations imposed by noisy quantum hardware on establishing long-distance end-to-end entanglements. Even considering a gate fidelity value of $f_\text{gate} = 0.9$, the maximum number of sequential swaps is limited to $3$ for $f_i \leq 0.9$, severely restricting the ability to establish long-distance end-to-end entanglements.

Similarly, the standard deviation of $f_\text{gate}$ across different repeaters significantly impacts the maximum achievable path length, as shown in \mbox{Fig.~\ref{fig:max_number_of_hops_std}} for $f_\text{gate}$ sampled from a $\min(1, \mathcal{N}(1, \sigma_{f_\text{gate}}^2))$. These plots were generated by analyzing $10000$ individual paths; the line shows the median maximum number of swaps, and the shaded area represents the $25^\text{th}$ to $75^\text{th}$ percentile range. For instance, with a standard deviation of $0.1$, the median maximum achievable path length for initial link fidelities of $0.8$ and $0.9$ drops to just $2$ and $5$ swaps, respectively.

\subsection{Hyperparameters}\label{subsec:hyperparams}
In Fig.~\ref{fig:hyperparams}, we show the influence of different hyperparameters on training of RELiQ.
RELiQ is robust to different exploration schedules and performs well even if very little exploration is allowed (compare Fig.~\mbox{\ref{fig:exploration}}).
However, the discount factor $\gamma$ significantly influences the performance of the trained model. While $\gamma = 0.9$ and $\gamma = 0.95$ produce similar results, $\gamma = 0.99$ leads to a decrease in performance.
This is caused by the reduced incentive of the model to choose shorter paths, leading to increased consumption of elementary links and ultimately to fewer end-to-end entanglements.
Similarly, the learning rate $lr$ significantly influences the training of RELiQ:
Training is stable for a learning rate of $10^{-4}$ or $5\cdot 10^{-3}$, but a learning rate of $10^{-3}$ leads to instability in training and drastic performance degradation.


\end{document}